\documentclass[a4paper,fleqn,usenatbib,useAMS]{mnras}

\usepackage{graphicx}	
\usepackage{amsmath}	
\usepackage{amssymb}	
\usepackage{multicol}        
\usepackage{bm}		
\usepackage{pdflscape}	

\usepackage[T1]{fontenc}
\usepackage{ae,aecompl}

\usepackage{txfonts}

\title[External FUV photoevaporation of discs]{Photochemical-dynamical models of externally FUV irradiated protoplanetary discs}
\author[T. J. Haworth et al.]
{\parbox{\textwidth}{Thomas J. Haworth$^{1, 2}$\thanks{E-mail: \texttt{thaworth@ast.cam.ac.uk}},
Douglas Boubert$^{1}$, Stefano Facchini$^{3}$, Thomas G. Bisbas$^{3,4}$ and Cathie J. Clarke$^{1}$ 
}\vspace{0.4cm}\\
\parbox{\textwidth}{$^{1}$Institute of Astronomy, Madingley Road, Cambridge, CB3 0HA, UK\\
$^{2}$  Astrophysics Group, Imperial College London, Blackett Laboratory, Prince
Consort Road, London SW7 2AZ, UK \\
$^{3}$ Max-Planck-Institut f\"ur Extraterrestrische Physik, Giessenbachstrasse 1, 85748 Garching, Germany\\
$^{4}$ Department of Astronomy, University of Florida, Gainesville, FL 32611, USA \\
}}

\begin{document}

\date{Accepted ???. Received ???; in original form ???}

\pagerange{\pageref{firstpage}--\pageref{lastpage}} \pubyear{2016}

\maketitle
\label{firstpage}

\begin{abstract}
There is growing theoretical and observational evidence that protoplanetary disc evolution may be significantly affected by the canonical levels of far ultraviolet (FUV) radiation found in a star forming environment, leading to substantial stripping of material from the disc outer edge even in the absence of nearby massive stars. In this paper we perform the first full radiation hydrodynamic simulations of the flow from the outer rim of protoplanetary discs externally irradiated by such intermediate strength FUV fields, including direct modelling of the photon dominated region (PDR) which is required to accurately compute the thermal properties. We find excellent agreement between our models and the semi--analytic models of \cite{2016MNRAS.457.3593F} for the profile of the flow itself, as well as the mass loss rate and location of their ``critical radius''. {This both validates their results (which differed significantly from prior semi--analytic estimates) and our new numerical method, the latter of which can now be applied to elements of the problem that the semi--analytic approaches are incapable of modelling}. We also obtain the composition of the flow, but given the simple geometry of our models we can only hint at some diagnostics for future observations of externally irradiated discs at this stage. We also discuss the potential for these models as benchmarks for future photochemical--dynamical codes. 
\end{abstract}

\begin{keywords}
accretion, accretion discs -- circumstellar matter -- protoplanetary discs --
hydrodynamics -- planetary systems: formation -- photodissociation region (PDR)

\end{keywords}

\section{introduction}
It is widely accepted that planets form from circumstellar discs of material found around young stars, composed of matter left over from the initial star formation process - so called protoplanetary discs. To understand the huge diversity of exoplanetary system architectures that are being discovered \citep{2015ARA&A..53..409W} we have to understand, not just the many physical mechanisms associated with the evolution of an isolated protoplanetary disc \citep[e.g.][]{2011ARA&A..49..195A, 2015arXiv150906382A}, but also how the evolution of discs varies as a function of environment. In particular, stars typically form in clusters, so gravitational encounters and radiation from other stars in the cluster are each likely to play a role. Although the former is expected at some level theoretically \citep[e.g.][]{2014MNRAS.441.2094R, 2015A&A...577A.115V}, only a small number of obvious interactions are currently observed to be taking place, for example the star-disc encounter in RW Aur \citep{2006A&A...452..897C, 2015MNRAS.449.1996D}. Furthermore, using N--body simulations \cite{2001MNRAS.325..449S}  concluded that the impact of dynamical interactions in Orion nebula cluster (ONC) like-environments is negligible compared to external photoevaporation. Obvious instances of radiative influence are the ``proplyd'' systems exposed to a very high energy radiation field from nearby O stars \citep[e.g.][]{1993ApJ...410..696O, 1996AJ....111.1977M, 1998AJ....116..322H, 2000AJ....119.2919B, 2001AJ....122.2662O, 2002ApJ...566..315H, 2003ApJ...587L.105S, 2012ApJ...746L..21W} which are relatively numerous compared to (obviously) gravitationally interacting star-disc systems.

\smallskip

Although we observe more proplyds than gravitational encounters, out of the total number of protoplanetary discs observed, only a very small fraction are sufficiently close to a sufficiently massive star to obviously be identified as a proplyd. For example \cite{2008AJ....136.2136R} find approximately 200 of  3200 stars in Orion are proplyds ($\sim6$ per cent). For the majority of star-disc systems, the local external radiation field is much weaker, but still of order tens to thousands of times stronger than the field measured in the immediate vicinity of the solar system. For example in \cite{2016arXiv160501773G} far ultraviolet (FUV) fields up to of order a few thousand G$_0$\footnote{Two measures of the FUV field are referred to in this paper. One is the Habing, which is $1.6\times10^{-3}$\,erg\,cm$^{-2}$\,s$^{-1}$, usually denoted G$_0$ and is a measure of the radiation field local to the solar system \citep{1968BAN....19..421H}. The other is the Draine, which is 1.71\,G$_0$ \citep{1978ApJS...36..595D}.} were estimated in the proximity of the Cygnus OB association. Furthermore this study found a reduction in the fraction of disc hosting stars in regions of higher UV flux that is inconsistent with either an age gradient or star-disc encounters.  It therefore seems likely that cluster radiation is directly shortening the lifetimes of protoplanetary discs in this system.  \cite{2016arXiv160701357M} also analysed $\sigma$ Ori, finding local FUV fields  in the range $300-1000$\,G$_0$ for 80 per cent of the discs they studied.  Additionally, important direct evidence of the impact of intermediate strength radiation fields on discs came from the identification of 7 proplyds within 0.3\,pc of the B1V type star HD\,37018 (42 Orionis) in Orion by \cite{2016ApJ...826L..15K}, where the FUV field in the vicinity of the proplyds was a maximum of  $\sim3000$\,G$_0$.

\smallskip

Protoplanetary discs are also irradiated internally by their host star. For example \cite{2014ApJ...784..127F} find typical UV fluxes at 1\,AU of $\sim$10$^7$\,G$_0$ for a sample of 16 T Tauri stars covering a range of evolutionary stages. Although the radiation from the host star suffers far less geometric dilution than that from other cluster members, the cluster radiation field is cumulative and, from some stars at least, will reach the disc surface and outer edge without having to traverse an optically thick disc. The magnitude of FUV fluxes from the central star reaching the disc outer surface is therefore exceeded by the external FUV flux. As a further example, \cite{2012A&A...541A..91B} ran radiative transfer and chemistry models of discs in hydrostatic equilibrium and found fields of less than 100\,G$_0$ throughout most of the disc mass and only a few G$_0$ at the disc outer edge, which is smaller than the external fields of order 1000\,G$_0$ found by  \cite{2016arXiv160501773G}, \cite{2016arXiv160701357M} and \cite{2016ApJ...826L..15K}.

\smallskip

Given the above, external irradiation is apparently pervasive in clusters and dominant at large disc radii over radiation from  the host star. The main effect of external irradiation is to drive an outflow from the disc outer edge (discussed in more detail in section \ref{sec:review}). This process is difficult to model because in the event that FUV irradiation dominates over EUV, the temperature and composition of the irradiated gas is not governed by photoionisation, but rather by photon dominated region (PDR) chemistry. PDR modelling is a highly complex procedure, involving many hundreds of species and thousands of reactions \citep[see e.g.][]{1985ApJ...291..722T, 1995ApJS...99..565S, 2015MNRAS.450.4424B}. It is therefore computationally expensive to compute the PDR conditions and doing so iteratively with a hydrodynamics solver is substantially more expensive. Analytic arguments and semi--analytic models have therefore been the only way to estimate the flow structure and mass loss rates \citep[reviewed in section 2, but some key papers are][]{1994ApJ...428..654H, 1998ApJ...499..758J, 1999ApJ...515..669S, 2004ApJ...611..360A, 2016MNRAS.457.3593F}.  In the latter two of these papers the approach is to pre-tabulate temperatures as a function of the local number density and radiation field (this latter depending on its unextincted value and the local extinction). Solving for the conditions at some point in the flow, applying this pre-tabulated grid and imposing conservation laws can thus yield the flow structure. These semi-analytic models are only valid within the confines of the pre--tabulated temperature grid (where interpolation is reasonable) and also only apply in steady state and (to date) spherical scenarios. Semi--analytic models also cannot be applied to arbitrary irradiated discs; only particular scenarios have solutions that can be computed semi--analytically. 

\smallskip

Until now, direct modelling of the PDR chemical and thermal properties in discs has been confined to hydrostatic (1+1D) scenarios \citep[e.g.][]{2008ApJ...683..287G, 2015ApJ...804...29G} and not to the flow
solutions expected in externally irradiated discs. However with advances in computing power and techniques it is finally becoming possible to directly couple PDR modelling to a full hydrodynamical framework \citep[a discussion of the current and near-future status of such modelling is given in][]{2016arXiv160801315H}. To date we are aware of two codes capable of this: one is that of \cite{2015ApJ...808...46M} which was constructed specifically with PDR-hydrodynamics in mind, but is presently confined to 2D. The other is  \textsc{torus-3dpdr} \citep[used here, see section \ref{numMeth}; ][]{ 2015MNRAS.448.3156H, 2015MNRAS.454.2828B} which is capable of 3D multiphysics, including  hydrodynamics, gravity, photoionisation, PDR chemistry and radiation pressure. We are just now gaining access to the tools that will allow us to directly model non-hydrostatic systems, a key example of which is externally irradiated protoplanetary discs. Given that such models are complicated and computationally expensive, it is crucial to ensure that the numerical implementation is robust before applying it to completely unexplored problems. 

In this paper we perform the first full radiation hydrodynamic simulations of (geometrically) simple externally irradiated protoplanetary discs. A key objective of our work is to validate our new numerical approach, which will be used in the future to perform simulations that are inaccessible semi--analytically (for example, models that probe regions of the parameter space that do not yield semi-analytic solutions, or 2D models that include multidimensional stratified discs).  In this paper we also aim to test the existing semi-analytic models. As we shall see in the next section \cite{2016MNRAS.457.3593F} predicts critical radii and mass loss rates that are different to \cite{2004ApJ...611..360A}. We can therefore verify whether these new solutions are accurate.

\section{Scenario overview/review of semi-analytic models}
\label{sec:review}
We begin by reviewing the historical models of external disc photoevaporation, which is essential in order to understand the key features (locations, time scales, length scales) of the process  required for dynamical modelling. 

According to prior studies \citep[e.g.][]{1994ApJ...428..654H, 1998ApJ...499..758J, 1999ApJ...515..669S} the nature of external radiatively driven mass loss depends strongly upon the disc size relative to the the radius at at which the gas sound speed equals the escape velocity 
\begin{equation}
	R_g = \frac{\mu m_H}{k_{B}T} G M_*.
\end{equation}
Previously this was predominantly referred to
as the critical radius but we here instead describe this as the
gravitational radius in order to distinguish it from the location of the
critical point of the flow which we define below \citep[see also][]{2016MNRAS.457.3593F}. The gravitational radius is typically slightly radially outward of the sonic point \citep[the point at which the flow transitions from the subsonic to supersonic regime, e.g. ][]{2004ApJ...611..360A}.  Photoevaporation is vigorous in the
case of discs larger than the gravitational radius, so this
regime was the focus of early photoevaporation studies \citep[e.g.][]{1994ApJ...428..654H, 1998ApJ...499..758J, 1999ApJ...515..669S}. In the region of supersonic
flow beyond the gravitational radius the flow velocity is a weak
function of radius and thus, under the assumption of spherical
symmetry and steady flow, the density distribution is of the form: 
\begin{equation}
	n(R) = n_g\left(\frac{R_g}{R}\right)^2
	\label{nbound}
\end{equation}
\citep{1998ApJ...499..758J} where $n_g$ is the number density at the gravitational radius. In the case that  the gravitational radius is interior to the disc outer edge, \cite{1998ApJ...499..758J} assume that the $R_g$ and $n_g$ terms are replaced by the disc outer radius $R_d$ and density at the disc outer radius $n_d$ respectively. 

An important development was made by \cite{2004ApJ...611..360A} who demonstrated that substantial flow rates can be driven from discs whose outer edges
are {\it interior} to the gravitational radius (down to radii
$\sim 0.1-0.2 R_g$). They argued that the mass loss problem can be considered as a 1D spherical scenario, since the mass loss is expected to be dominated by material stripped from the outer disc edge (see the appendix of their paper). This expectation arises for two reasons. Firstly the vertical scale height is much smaller than the radial, which \cite{2004ApJ...611..360A} suggest results in a higher density at the radial sonic point  than the vertical and therefore a higher  radial mass loss rate. The second argument is that material at the disc outer edge is less gravitationally bound to the system than most of the material being photoevaporated vertically. They found that the flow can be described in terms of a spherical
wind that is analogous to a Parker wind \citep{1965SSRv....4..666P} but with
a non-isothermal structure and with the inclusion of centrifugal
terms associated with the finite angular momentum of the
evaporating disc material. This allowed them to semi-analytically calculate steady state mass loss profiles from discs with outer radii ($R_d$) interior to the gravitational radius, solving the streamline equation radially outwards from the disc outer edge to the sonic point in the flow. They computed the temperature as a function of column, local FUV field and local number density in their models using a lookup table generated by the PDR modelling code of \cite{1999ApJ...527..795K}. 

\cite{2016MNRAS.457.3593F} {(hereafter, F16)} also addressed the problem using semi-analytic models, but took a slightly different approach, in that they located a critical point in the modified Parker wind solution such that, at this point, the velocity
gradient is given by the ratio of two terms that are each individually
zero. This leads to the definition of a critical flow velocity that is
\begin{equation}
	u_c^2 = f + g \frac{\partial f}{\partial g}
	\label{equn:vcrit}
\end{equation}
where $f$ is the dimensionless temperature $T/T_c$ and $g$ the dimensionless density $n/n_c$ and subscript $c$ terms are quantities at the critical point. A cylindrical version of this spherical result is also derived in Appendix \ref{cylsol} of this paper, which illustrates the approach and steps involved in more detail. {F16} solved for the properties at this critical point for a given value of $T_c$ (using the PDR model to evaluate the non-isothermal terms
such as the second term in equation \ref{equn:vcrit}); these non-isothermal terms turned out
to be important for determining the mass loss rates. With the conditions at the critical point known, they then integrated from that point inwards to the disc outer edge using the standard conservation equations \citep[conversely][integrated from the disc outer edge outwards to the sonic point]{2004ApJ...611..360A}. Because these solutions first solve for the conditions at the critical point {F16} were able to compute solutions over a wider parameter space than previously possible (e.g. for low radiation field strengths and larger outer disc radii). They also found that their critical point could be larger than the sonic radius by a factor of a few and that their mass loss rates were lower than those computed by \cite{2004ApJ...611..360A} (though this could in part be due to the different PDR codes employed). {F16} also accounted for the fact that of the dust, which sets the opacity in the flow, only small grains are entrained in the flow \citep[e.g.][]{2016MNRAS.461..742H} and so dust to gas mass ratios can be very low, especially if there is significant grain growth in the disc (e.g. $10^{-5}$ compared to the canonical $10^{-2}$ in the interstellar medium). This leads to reduced extinction and therefore more effective photoevaporation than expected when not accounting for grain growth. 

In summary, the models to date compute the steady state flow structure from the disc outer edge to either the sonic point  \citep{2004ApJ...611..360A} or critical radius (F16),  both assuming that the density structure at larger radius obeys an inverse  square law (equation \ref{nbound}).  There is one final key feature of this scenario. We distinguish between ``optically thick'' and ``optically thin'' regimes depending on whether the disc outer edge is optically thick or thin to the incident FUV irradiation. 


\section{Numerical method}
\label{numMeth}
We use the \textsc{torus-3pdr} code to perform the simulations in this paper \citep{2015MNRAS.454.2828B}. This code is the direct incorporation of algorithmic components and routines from the \textsc{3d-pdr} code \citep{2012MNRAS.427.2100B} into the Monte Carlo radiation transport and hydrodynamics code \textsc{torus} \citep{2000MNRAS.315..722H, 2012MNRAS.420..562H, 2015MNRAS.448.3156H}. The code is described in detail in the aforementioned papers; however in order to render the models in this paper more easily reproducible we  briefly summarise the main hydrodynamic features and reiterate the nature of the PDR chemistry in more detail.

\subsection{Radiation hydrodynamics}
\textsc{torus}(\textsc{-3dpdr}) performs radiation hydrodynamics simulations via operator splitting, e.g. computing radiative transfer, composition and thermal calculations in sequence with hydrodynamics updates \citep[][]{2015MNRAS.453.2277H}. The temperatures calculated from the radiative transfer/composition calculations then provide the gas pressure in the dynamical part of the calculation. In this paper, only hydrodynamic and PDR calculations are performed sequentially  - we do not include any of the other components or physics modules (e.g. radiation pressure) in \textsc{torus}. In this paper,  we assume that radiative processes dominate the temperature determination, and therefore do not include processes such as shock heating. 

\textsc{torus} uses a finite volume hydrodynamics algorithm that is total variation diminishing, uses Rhie-Chow interpolation \citep{1983AIAAJ..21.1525R} and, in this paper, uses a van Leer flux limiter \citep{vanleer}. 
The discs that we consider are relatively low mass, so we do not consider self-gravity of the gas, but do include a point source potential from the star hosting the disc, which is always assumed to be of mass 1\,M$_{\odot}$.

\subsection{Photochemistry and ``radiative transfer''}
{The photochemical component of the calculation is based on routines from the \textsc{3d-pdr} code \citep{2012MNRAS.427.2100B}, which in turn is based on the \textsc{ucl{\textunderscore}pdr} code \citep{2005MNRAS.357..961B, 2006MNRAS.371.1865B}}. In this paper we use essentially the same PDR reaction network used to construct the temperature grid by {F16}. That is, we use a reduced version of the \textsc{umist} 2012 network which consists of 33 gas species (including electrons) and 330 reactions \citep{2013A&A...550A..36M}. The species and initial abundances considered are summarised in Table \ref{PDRGuts}, where the sum of the hydrogen abundances equals unity ($X(H) + 2X(H_2) = 1$) and the other abundances are relative to atomic hydrogen. We include the modifications to the base \textsc{umist} network that are discussed in section 2 of \cite{2014MNRAS.443..111B}, excluding their comments on dust properties (which we will discuss below). Note that using the full \textsc{umist} network leads to differences in the temperature of up to about 10 per cent, so the dynamics is quite accurately modelled with the reduced network for a significantly reduced computational cost.

\begin{table}
 \centering
  \caption{A summary of the species included and initial gas abundances for the reduced network used in this paper. The sum of  hydrogen atoms in atomic and molecular hydrogen is unity. The other abundances are with respect to atomic hydrogen.}
  \label{PDRGuts}
  \begin{tabular}{@{}l l@{}}
  \hline
 Gas species with non-zero & H ($4\times10^{-1}$), H$_2$ ($3\times10^{-1}$), \\
  	initial abundance & He ($8.5\times10^{-2}$), C+ ($2.692\times10^{-4}$) \\
	&O ($4.898\times10^{-4}$), Mg+ ($3.981\times10^{-5}$) \\
	& \\	
   Other gas species & e$^-$, H+, H$_2$+, H$_3$+, He+, O+, O$_2$, O$_2$+, \\
   			  &  OH+, C, CO, CO+, OH, HCO+, Mg, H$_2$O,  \\
			  &  H$_2$O+, H$_3$O, CH, CH+, CH$_2$, CH$_2$+, \\
			  &  CH$_3$, CH$_3$+, CH$_4$, CH$_4$+, CH$_5$+, \\
			& \\
   Other species & Cosmic rays, PAHs, Dust\\
\hline
\end{tabular}
\end{table}

In this paper we are comparing dynamical models with semi-analytic solutions in relatively simple geometries. We therefore do not require the sophisticated radiation transport capabilities available in \textsc{torus}. Rather, we simply specify a ``UV'' field (that integrated from 912-2400\,\AA) at infinity, in Draines \citep[where 1 Draine = $1.71$\,G$_0$;][]{1978ApJS...36..595D}. This radiation field is attenuated along a single ray from infinity down to the disc edge. Note that geometric dilution is \textit{not} included; the radiation field is only attenuated through absorption (the light is a beam). The opacity of the medium is a crucial parameter, which {F16} set based on the mean grain cross section $\sigma_{\textsc{FUV}}$. The radiation (UV) field at a given point is thus
\begin{equation}
	\chi = \chi_o\exp\left({-\frac{\sigma_{FUV}}{1.8}N\tau_{FUV}}\right). 
	\label{chiEQN}
\end{equation}
where $\chi_o$ is the radiation field at infinity, $N$ is the column density from infinity to the point at which the UV field is being evaluated and $\tau_{FUV}=3.02$ is a parameter converting from extinction to UV attenuation. The extinction from infinity to a given point in the flow is thus
\begin{equation}
	A_V = N\frac{\sigma_{FUV}}{1.8}. 
	\label{AVEQN}
\end{equation}
{F16} directly calculated the mean grain cross section; however in this paper where we are primarily trying to validate the semi-analytic solutions, we use cross sections appropriate for a given semi-analytic model.  Note that the dust grains are assumed to be spherical and that the dust to gas mass ratio is always assumed to be $10^{-5}$ \textit{in the flow}. This low dust to gas mass ratio arises because only small grains are entrained in the flow; grain growth in the disc therefore depletes the flow of dust (F16). \textsc{torus-3dpdr} uses a single representative grain radius for the PDR calculation; the total surface area of grains is the important quantity rather than the size distribution.  We find that at the low dust to gas ratio in the flow the result is rather chemically and thermally insensitive to this representative grain radius (i.e. we get similar results for a representative grain radius of 1\,mm or 0.1\,$\mu$m, though this does not hold for the canonical dust to gas ratio of $10^{-2}$ in the interstellar medium or disc).  From {F16},  typically grains entrained in the flow are expected to be $\sim0.1\mu$m.  We assume chemical equilibrium, which means that the timescales for chemical reactions are not limited by the duration of the hydrodynamic time steps (a discussion of timescales is given in section \ref{sec:Timescales}).

An escape probability estimate is required to compute the PDR gas level populations and ultimately therefore the temperature/abundances. \textsc{torus-3dpdr} makes this estimate by casting $\textsc{healpix}$ rays to sample $4\pi$ steradians; however in {F16} only escape along the radial direction is considered (i.e. a single ray traced radially outward away from the disc is used for the escape probability estimate, with other directions assumed to be infinitely optically thick). We refer to their approach as using \textit{radially dominated photon escape}.  In this paper we are limiting ourselves to dynamical models analogous to past semi-analytic studies, so also retain the use of a radially dominated escape probability approximation. We find that in this scenario the heating is insensitive to the assumed microturbulent velocity\footnote{{F16} used a supersonic turbulent velocity of 1.5\,km/s (the default value in the PDR code they used). We ran most of our models with this same value to remain consistent, but ran test models to verify that more realistic subsonic turbulent velocities gave a negligibly different result.}.

\subsubsection{Heating, cooling and temperature determination}
The heating and cooling mechanisms are again the same as those included in the PDR lookup tables by {F16} and the interested reader should refer to Figure 2 in that paper. In summary though, 
heating is dominated by photoelectric heating from atomic layers of polycyclic aromatic hydrocarbons (PAHs). PAHs are hydrodynamically well coupled to the gas and there is no clear evidence of how dust growth affects their abundance. The PAH-to-gas ratio is therefore assumed to not be depleted in the flow. We follow the aforementioned authors in assuming a PAH to dust mass ratio of $2.6\times10^{-4}$. Other heating contributions are C ionisation, H$_2$ formation and photodissociation, FUV pumping, cosmic rays, turbulent and chemical heating and gas-grain collisions. The cooling contributions are dominated by lines of CO, O\,I, C\,I, C\,II and again gas-grain interactions, with O\,I and C\,II dominating in optically thin regions. We account for heating from the disc's host star in the flow by prescribing the maximum of the temperature calculated by the PDR code and that given by 
\begin{equation}
	T_{*} = \left(\frac{100}{\textrm{K}}\right)\left(\frac{R}{\textrm{AU}}\right)^{-1/2}.
	\label{tBound}
\end{equation}
The temperature in the disc is always assumed to be set by equation \ref{tBound}. Again, this is not always going to be the most realistic prescription (in particular, where the disc outer edge is optically thin to the incident FUV field) but is consistent with past semi-analytic models.

In this paper we also assume thermal equilibrium, which we checked was the case by comparing the thermal and dynamical timescales in our actual simulations once they had attained a steady state (see section \ref{sec:Timescales}). That is, at least once a steady state is attained, we find thermal equilibrium to be valid. In the PDR phase of our simulations chemical and thermal equilibrium are solved iteratively until convergence, as detailed in \cite{2012MNRAS.427.2100B, 2015MNRAS.454.2828B}. As already mentioned, PDR and hydrodynamical components of the simulations are performed iteratively. After the first PDR calculation, the initial temperature and composition guesses for subsequent PDR calculations are the previously computed values: this means that later PDR calculations start closer to convergence (especially once the system is $\sim$steady) and therefore run much more quickly than the initial step.

\subsection{Boundary conditions}
\label{bounds}
Although we include discs in our simulation grids, we do not allow them to evolve. The inner boundary of the flow is the disc outer edge. We construct discs by imposing pressure equilibrium between the innermost part of the flow (from a given semi-analytic model) and the disc outer edge, assuming that the disc outer temperature is given by 
\begin{equation}
	T_{\textrm{outer}} = \textrm{max}\left(100\left(\frac{R_{\textrm{d}}}{\textrm{AU}}\right)^{-1/2}, 10\right)\,\textrm{K}
	\label{tBound2}
\end{equation}
where $R_d$ is the disc outer radius. With the temperature known, pressure equilibrium thus yields the disc outer mid plane density. We assume the disc temperature profile follows equation \ref{tBound} and the disc radial surface density profile is of the form $\Sigma = \Sigma_d(R/R_d)^{-1}$. Using the conditions defined at the disc outer edge therefore yields a disc mass via
\begin{equation}
	M_d=2\pi R_d^2 \Sigma_d
\end{equation}
where $R_d$ is the disc outer radius and $\Sigma_d$ is the surface density at the disc outer edge. We do quote our inner boundary condition properties for each model in this paper in tabulated parameter summaries (Tables \ref{modelSummaryA}-\ref{modelSummaryB}). Note that the inner boundary between the disc and flow can be a contact discontinuity in density and temperature. We also impose Keplerian rotation throughout the ``disc'' component of the simulation. We do not allow the temperature, density or velocity in the disc region to evolve away from the prescription detailed here. 

There are two options regarding the outer boundary. One possibility is to model the flow out to the critical radius only, akin to the semi-analytic models. In this case an outer boundary extinction to infinity needs to be imposed; following \cite{1998ApJ...499..758J} we assume that the density beyond the critical radius is of the form 
\begin{equation}
	n(R>R_c)=n_c \left(\frac{R_c}{R}\right)^2
	\label{nbound2}
\end{equation}
i.e. the column density from the critical radius (which in this case would be the grid outer edge) to infinity is just the density at the critical radius times the critical radius. Equation \ref{AVEQN} then gives the boundary extinction.  Typically this boundary extinction is very small over a substantial fraction of the flow from the disc outer edge to the critical radius. We found this approach could give somewhat unstable results because the nature of the flow is sensitive to the critical radius, which might not be properly captured using this technique. The other alternative is to use a model grid substantially  beyond the critical radius, in which case we impose no boundary extinction since it becomes even more negligible. For the models presented in this paper we employ the latter approach, but in our model specifications include the relevant information required to use the former approach should someone so desire. We employ a free outflow/no inflow hydrodynamic outer boundary condition.

\subsection{Parallelisation, expense and optimisation}
\textsc{torus} is highly optimised and hybrid \textsc{openmp}--\textsc{mpi} parallelised, including domain decomposition. Many of these features are retained in \textsc{torus-3dpdr}, but some of the larger scale optimisations, such as a \textsc{healpix} analogue of the ``multiple grid copies'' technique for Monte Carlo radiation transport described by \cite{2015MNRAS.448.3156H}, are not yet fully implemented.  The simulations in this paper therefore run on only small numbers of cores (i.e. 3 for 1D models and 5 or 17 for 2D models  - numbers governed by the domain decomposition scheme). Fortunately with hindsight from our initial (more expensive) modelling efforts we know that the simulations in this paper do indeed fall into a steady state. The computational expense can therefore be drastically reduced by performing PDR temperature updates only periodically. For example, all models in this paper give equally accurate \textit{steady state} results if the PDR step is performed every single or every many thousand hydrodynamic steps (e.g. we ran a model where the PDR update took place every $\sim4$\,kyr and still found good agreement). However the way in which this steady state is reached will vary depending on how frequently PDR calculations are performed. A further optimisation already mentioned is that in subsequent PDR steps, by starting with the prior temperature/abundances computed, once the simulation approaches a steady dynamical state most cells will begin new PDR steps close to convergence. Given that in our 1D spherical models with a maximum number of 1024 cells the PDR steps can take $\sim$1\,hour (on 3 Intel Xeon E5 2.4GHz processors) and the hydrodynamics steps $\sim$0.1\,s the value of these optimisations can't be understated. We note that in more realistic future applications the expense will be even greater given that the radially dominated photon escape approximation should be droppend and the 3D escape probabilities computed. Furthermore, in future applications the radiation field should also be directly computed rather than being a single beam, immune to geometric dilution. Lastly, if the time evolution of the system is also of interest (e.g. it is not entirely steady) then PDR steps will likely have to be taken much more frequently (see section \ref{sec:Timescales}).

\begin{figure}
	\hspace{-15pt}
	\includegraphics[width=6.6cm, angle=270]{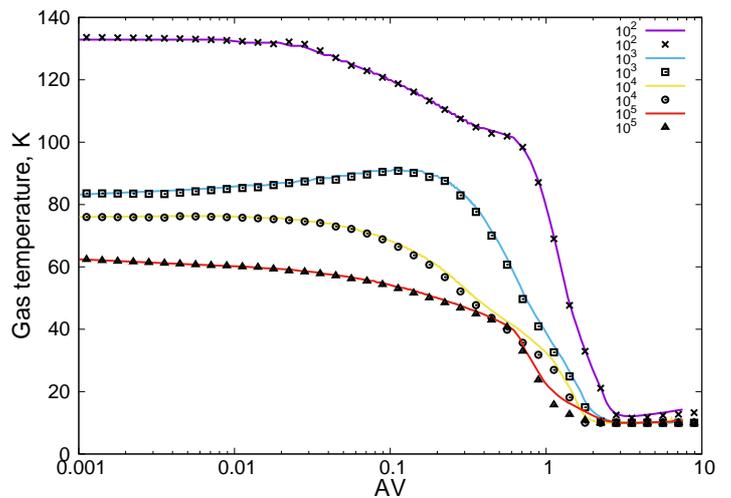}
	\caption{Slabs of different number density being irradiated by a 30 Draine field. The solid lines are computed by \textsc{torus-3dpdr}, the code used in this paper. The points denote the result from F16, which is reassuringly consistent given that the chemical networks etc. should be very similar.}
	\label{slabTempFig}
\end{figure}

\section{Code validation}
\label{validCode}
The hydrodynamics/radiation hydrodynamics in \textsc{torus} has been validated in \cite{2012MNRAS.420..562H} and \cite{2015MNRAS.453.1324B}.  The PDR components of \textsc{torus-3dpdr} have also been validated in \cite{2015MNRAS.454.2828B} which compared against results consistent with the \cite{2007A&A...467..187R} tests.  Before running any dynamical simulations, we verified that our code computes temperatures as a function of extinction consistent with that used by {F16} to generate lookup tables for the semi-analytic solution. An example is given in Figure 
\ref{slabTempFig}, where a 30 Draine (not Habing) field irradiates a medium of number density 100\,cm$^{-3}$, a grain radius of 1\,$\mu$m, dust to gas ratio of $10^{-5}$ and  the default \textsc{3d-pdr} $\sigma_{FUV}=1.13202\times10^{-21}$\,cm$^2$.  This is analogous to the top left panel of Figure 3 from  {F16}. Note that these are for \textit{planar} slabs (i.e. cartesian, not spherical) in the single-ray limit. That is, the gas is assumed to be infinitely optically thick in all directions other than that from which the radiation is impinging.

\begin{table*}
 \centering
  \caption{A summary of the parameters used for the spherical models in this paper. SA denotes a semi-analytic parameter which is not required to run the model (unless modelling only out to the critical radius), but is included for reference. }
  \label{modelSummaryA}
  \begin{tabular}{@{}l l l l l l@{}}
  \hline
  Parameter & Model A & Model B & Model C & Model D & Description \\
  \hline
  \hline
  Geometry & Spherical & Spherical & Spherical & Spherical &  \\
   $M_d$ (M$_{\odot}$) &  $3.7\times10^{-3}$  & $1.1\times10^{-5}$  & $1.42\times10^{-3}$ & $1.64\times10^{-6}$  & Disc mass\\       
   $R_d$ (AU) & 180  & 180  & 80 & 30 &  Disc outer radius \\  
   $T_d$ (K) & 10.0 & 10.0  & 11.18  & 18.26  & Temperature at disc outer edge\\   
   $\rho_d$ (g\,cm$^{-3}$) & $5.28\times10^{-16}$   & $1.55\times10^{-18}$   & $3.27\times10^{-15}$  & $9.17\times10^{-17}$  & Density at disc outer edge \\   
   $H_d$ (AU) & 20.4  & 20.4  & 6.4 & 1.9  & Scale height at disc outer edge \\   
  FUV (Draines) & 30 & 30  & 300  & 3000 & FUV field at infinity \\  
  $M_*$ (M$_\odot$) & 1 & 1 & 1 & 1 & Host star mass \\
  $\rho_a$ (g\,cm$^{-3}$) & $1.67\times10^{-21}$ & $1.67\times10^{-21}$ & $1.67\times10^{-21}$ & $1.67\times10^{-21}$ &  Initial density external to disc \\   
  $\mu$ & 1.3 & 1.3 & 1.3 & 1.3 & Mean particle mass everywhere \\  
  $\sigma_{\textrm{FUV}}$ & $5.2674\times10^{-23}$  & $4.49809\times10^{-23}$  &  $5.11038\times10^{-23}$  & $2.81383\times10^{-23}$  &   FUV cross section \\  
  $D/G$ & $10^{-5}$ & $10^{-5}$ & $10^{-5}$ & $10^{-5}$ &  Dust to gas mass ratio \\     
  $R_c$ (AU) & 1180.85  & 785.4  & 511.75  & 257.6 &  Critical radius (SA) \\
  $\rho_c$  (g\,cm$^{-3}$ ) & $3.56\times10^{-20}$  &  $2.25\times10^{-21}$ & $2.50\times10^{-20}$  & $1.40\times10^{-21}$  &  Density at critical radius (SA) \\
  $N_B$ (cm$^{-2}$) & $5.2\times10^{20}$  & $2.2\times10^{19}$ & $1.6\times10^{20}$  & $4.5\times10^{18}$ &  Column to infinity from  $R_c$ (SA) \\
  $AV_B$ &  $1.525\times10^{-2}$  & $5.486\times10^{-4}$ & $4.439\times10^{-3}$ & $6.97\times10^{-5}$ &  Extinction to infinity from  $R_c$ (SA) \\
  $T_c$ (K) & 56  & 78  & 136 & 268 &  Temperature at critical radius (SA) \\  
  $c_c$  (km/s) &  0.6 & 0.7  & 0.9 & 1.3 &  Sound speed at critical radius (SA) \\
  $\tau_d$ & 5.76  & $1.04\times10^{-2}$  & 9.94 & $1.94\times10^{-2}$  & Optical depth to disc from infinity (SA) \\
 $\dot{M}$ (M$_{\odot}$\,yr$^{-1}$) & $1.42\times10^{-8}$& $9.55\times10^{-10}$ & $1.96\times10^{-9}$ & $1.08\times10^{-10}$ & Mass loss rate (SA)\\
\hline
\end{tabular}
\end{table*}

\bigskip

\begin{table*}
 \centering
  \caption{A summary of the parameters used for the cylindrical models in this paper. SA denotes a semi-analytic parameter which is not required to run the model (unless modelling only out to the critical radius), but is included for reference. }
  \label{modelSummaryB}
  \begin{tabular}{@{}l l l l@{}}
  \hline
  Parameter & Model E & Model F &  Description \\
  \hline
  \hline
  Geometry &  Cylindrical & Cylindrical &  \\
   $M_d$ (M$_{\odot}$) & $1.54\times10^{-5}$ & $1.65\times10^{-6}$ & Disc mass\\       
   $R_d$ (AU)  & 60 & 100 &  Disc outer radius \\  
   $T_d$ (K) &  12.9 & 10.0 &  Temperature at disc outer edge\\   
   $\rho_d$ (g\,cm$^{-3}$) &  $8.97\times10^{-17}$ & $1.8\times10^{-18}$ & Density at disc outer edge \\   
   $H_d$ (AU) & 4.5 & 8.5 & Scale height at disc outer edge \\   
  FUV (Draines) &  300  & 300 & FUV field at infinity \\  
  $M_*$ (M$_\odot$) & 1 & 1 & Host star mass \\
  $\rho_a$ (g\,cm$^{-3}$) &  $1.67\times10^{-21}$ & $1.67\times10^{-21}$ &  Initial density external to disc \\   
  $\mu$ &  1.3 & 1.3 & Mean particle mass everywhere \\  
  $\sigma_{\textrm{FUV}}$  & $3.80612475\times10^{-23}$ & $4.53065125\times10^{-23}$ &   FUV cross section \\  
  $D/G$  & $10^{-5}$ & $10^{-5}$ &  Dust to gas mass ratio \\     
  $R_c$ (AU)  & 885.75 & 890 &  Critical radius (SA) \\
  $\rho_c$  (g\,cm$^{-3}$ )   & $4.54\times10^{-22}$ & $1.25\times10^{-21}$ &  Density at critical radius (SA) \\
  $N_B$ (cm$^{-2}$)  & $2.77\times10^{18}$ & $7.67\times10^{18}$ &  Column to infinity from  $R_c$ (SA) \\
  $AV_B$ &  $5.86\times10^{-5}$ & $1.93\times10^{-4}$ &  Extinction to infinity from  $R_c$ (SA) \\
  $T_c$ (K) & 190 & 160 &  Temperature at critical radius (SA) \\  
  $c_c$  (km/s)   & 1.1 & 1 &  Sound speed at critical radius (SA) \\
  $\tau_d$  & $8\times10^{-2}$ & $4.3\times10^{-3}$ & Optical depth to disc from infinity (SA) \\
 $\dot{M}$ (M$_{\odot}$\,yr$^{-1}$) & $3.9\times10^{-12}$ & $2.0\times10^{-11}$ & Mass loss rate (SA)\\
\hline
\end{tabular}
\end{table*}

\section{Results and discussion}
We now move on to discuss photochemical-dynamical models of protoplanetary discs that are externally irradiated by an FUV radiation field. We first discuss the model parameters in section \ref{sec:modelParams}. We then check the dynamical and thermal timescales in the models in section \ref{sec:Timescales} which govern the simulation run time and whether the assumption of thermal equilibrium is prudent.  We then compare the dynamical models with semi-analytic solutions in \ref{sec:comparison}. Finally,  in section \ref{sec:flowComp}, we comment on the composition of the photoevaporative flow.

\subsection{Model Parameters}
\label{sec:modelParams}
We present a total of six models in this paper. Four of these (models A--D) are in a spherical geometry, akin to that considered by  \cite{2004ApJ...611..360A} and  {F16}. The other two (models E--F) use a cylindrical grid, the semi-analytic analysis for which is provided here in Appendix \ref{cylsol}, which will be useful for testing simulations in 2D geometries. A thorough list of the parameters associated with the spherical and cylindrical models is given in Tables \ref{modelSummaryA} and \ref{modelSummaryB} respectively. Note that the disc masses provided in these tables assume an $R^{-1}$ surface density profile and the temperature profile given by equation \ref{tBound}.  Since we are comparing with semi-analytical models, we quote a large number of parameters related to the semi-analytic result that are not generally required to set up an arbitrary dynamical simulation (such as the critical point and the conditions there). We include these values for reference and to provide the information necessary to treat the outer boundary according to equation \ref{nbound2} (if desired). Our choice of models represents a variety of disc outer radii (30, 60, 80, 100, 180\,AU), incident FUV fields (30, 300, 3000\,Draines) and optical depths to the disc outer edge. The optical depth is particularly important, since if the wind flow close to the disc outer edge is optically thick then the incident FUV field will have drastically decreased somewhere exterior to the disc outer edge. Two of our spherical models are optically thick (A and C) and two are optically thin (B and D). In addition to the parameters required to reproduce our models, we also include the expected photoevaporative mass loss rates in Tables  \ref{modelSummaryA}/\ref{modelSummaryB} which can be used by those wishing to reproduce these models to give a quick measure of the agreement between the semi-analytic models and their results. Furthermore, to easily permit more thorough comparison, we include the semi--analytic solutions for each model {(the solutions for models A--D are from F16 and and those for models E, F from the appendix of this paper)} as supplementary online data to the journal article. 

For the spherical models, which we run on a 1D grid, we use an adaptive grid up to a maximum effective number of cells of 1024, which provides accurate results at low computational cost (we have run models at higher and lower resolution and will discuss the potential issues with low resolution models in section \ref{sec:comparison}). We impose a steady `disc' structure
interior to the fixed outer dics's radius using the procedure discussed in section \ref{bounds}. Initially, beyond the disc outer edge we set a uniform density medium of $1.67\times10^{-21}$\,g\,cm$^{-3}$ ($\sim10^3$\,m$_{\textrm{H}}$). 

Our cylindrical scenarios are modeled on a two dimensional grid (\textsc{torus-3dpdr} does not currently support a 1D cylindrical geometry). Since the primary purpose of these models is to test the semi-analytic solutions, and high resolution models would be computationally expensive, we therefore run the cylindrical models using $128^2$ cells. Owing to the small number of cells in the disc and the fact that properties in the disc vary quite rapidly as a function of radius (which may lead to inaccurate boundary conditions on low resolution grids, section  \ref{sec:comparison}) we therefore  impose a flat ``disc'' with uniform properties that put the disc in pressure equilibrium with the innermost part of the flow. Given the resolution limitations on our cylindrical models, we (at this stage) only study them in a dynamical context, rather than commenting on the composition.

\subsection{Timescales}
\label{sec:Timescales}
There are two key timescales {that we} consider for these models. The first is a dynamical timescale, which we take as the time for material to flow from the disc outer edge to the critical point. This dynamical timescale should inform the run time required to reach a steady state solution.  The second is the thermal timescale $k_BT/\dot{Q}$, where $\dot{Q}$ is the heating rate per unit mass, which tells us whether the assumption of thermal equilibrium is valid depending on the thermal timescale relative to the flow timescale.  

\begin{figure}
	\hspace{-10pt}
	\includegraphics[width=6.5cm, angle=270]{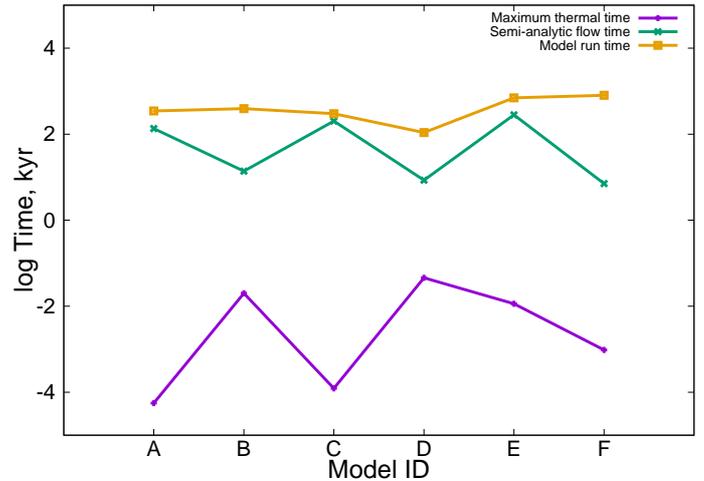}
	\caption{A summary of key timescales for the models in this paper. The purple line-points are the maximum thermal timescale in the flow. This timescale is always orders of magnitude lower than the flow timescale (from the disc outer edge to the critical point) which is given by the green line. The assumption of thermal equilibrium is therefore valid.  The orange line denotes the final simulation times of the models in this paper, which is in excess of the flow time.   }
	\label{fig:Timescales}
\end{figure}

Figure \ref{fig:Timescales} provides an overview of the key timescales, compared with the simulation run times. The green line-points represent the dynamical time, which is computed using the semi--analytic solution and ranges from $\sim$10\,kyr to $\sim$300\,kyr. The longer flow timescales are associated with models where the inner parts of the flow are optically thick to the incident radiation field, meaning that the temperature and induced velocities are lower.  The thermal timescale can only be computed using simulation results (where temperatures and heating rates have been computed - the latter is not included in the semi-analytic models unless they are further postprocessed using a full PDR calculation). The purple line-points in Figure \ref{fig:Timescales} represents the maximum thermal timescale for all models (the maximum, since the thermal timescale varies throughout the flow), which is always orders of magnitude lower than the dynamical time.  The assumption of thermal equilibrium is therefore prudent.

The orange line-points in Figure \ref{fig:Timescales} shows the simulation end times of all of the models in this paper, which are in excess of the dynamical time.  The models should therefore be in a steady state (we also verified this by directly checking the time evolution of the properties at the disc outer radius, critical point and grid outer radius). We note that the models typically appeared to achieve a steady state long before the flow timescale. 

{There is another fundamentally important timescale, which is that over which the FUV flux varies as the star-disc system evolves dynamically within the cluster. In the absence of highly perturbing gravitational encounters, which might drastically alter the FUV flux incident on a disc over a short timescale, this is given by the orbital timescale. \cite{2011PASP..123...14H} studied the time evolution of the FUV flux incident upon discs orbiting a cluster approximated by a Hernquist potential \citep{1990ApJ...356..359H} and found typical orbital periods of $\sim0.1-1$\,Myr. The orbital timescale in such a model is therefore of order our flow timescale. Although we reiterate that steady state is  typically achieved in our models much more rapidly than the flow timescale, it is important to note that in reality the propagation of the star-disc through the cluster may compromise the assumption of steady state. Non-steady models could be studied in the future using our numerical approach.}

\begin{figure*}
	\hspace{-10pt}
	\includegraphics[width=6.25cm, angle=270]{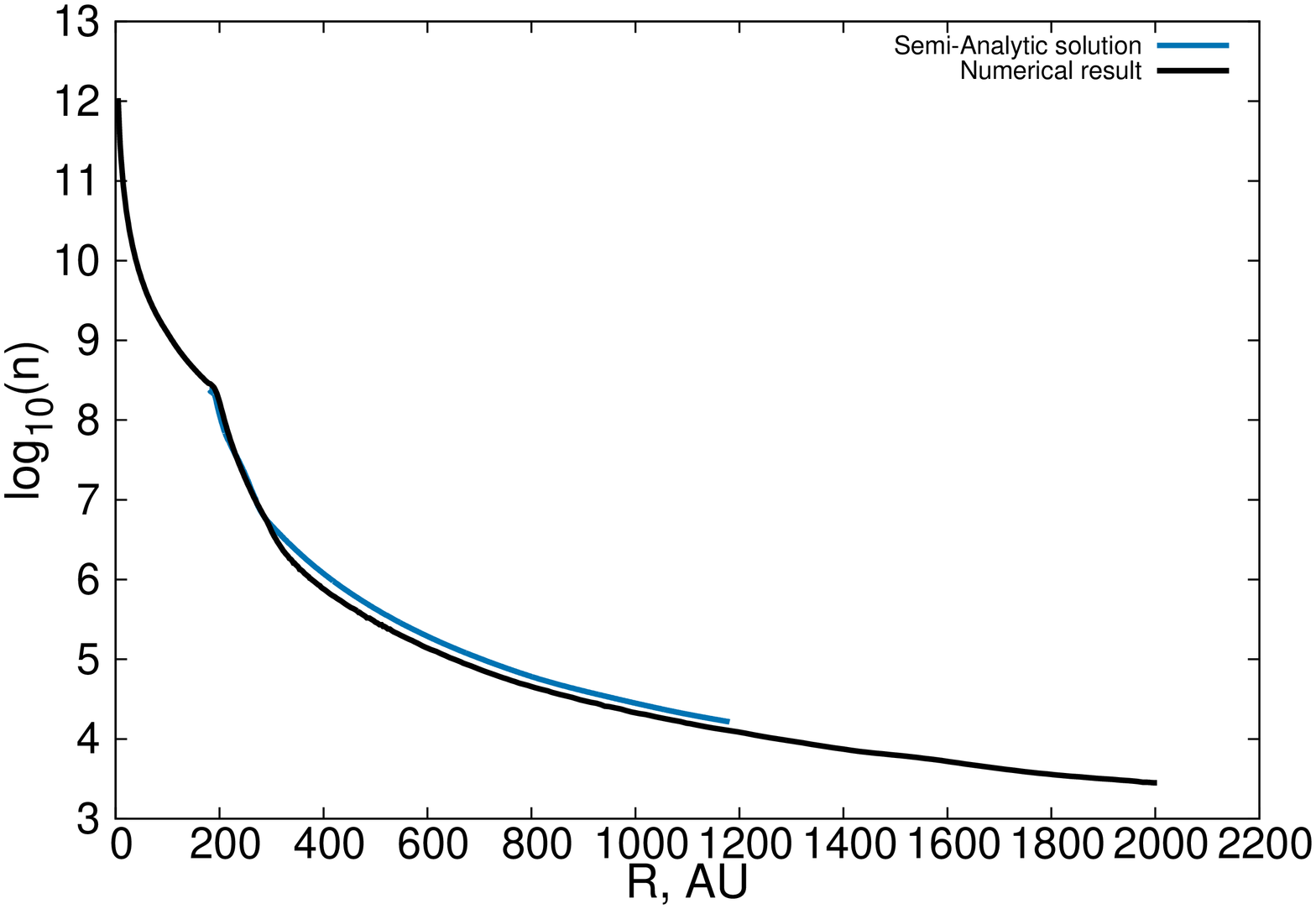}	
	\includegraphics[width=6.25cm, angle=270]{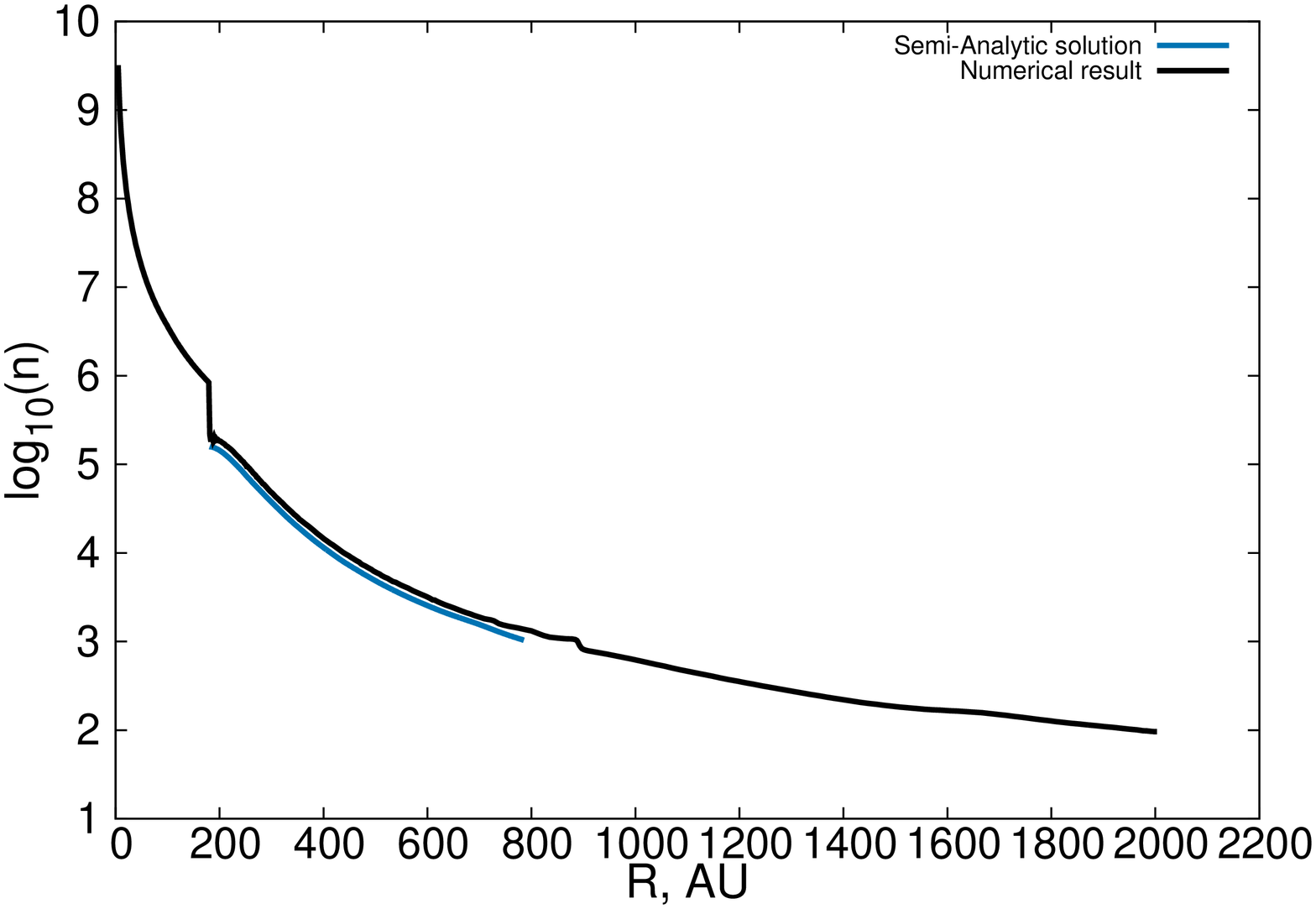}		

	\hspace{-10pt}
	\includegraphics[width=6.25cm, angle=270]{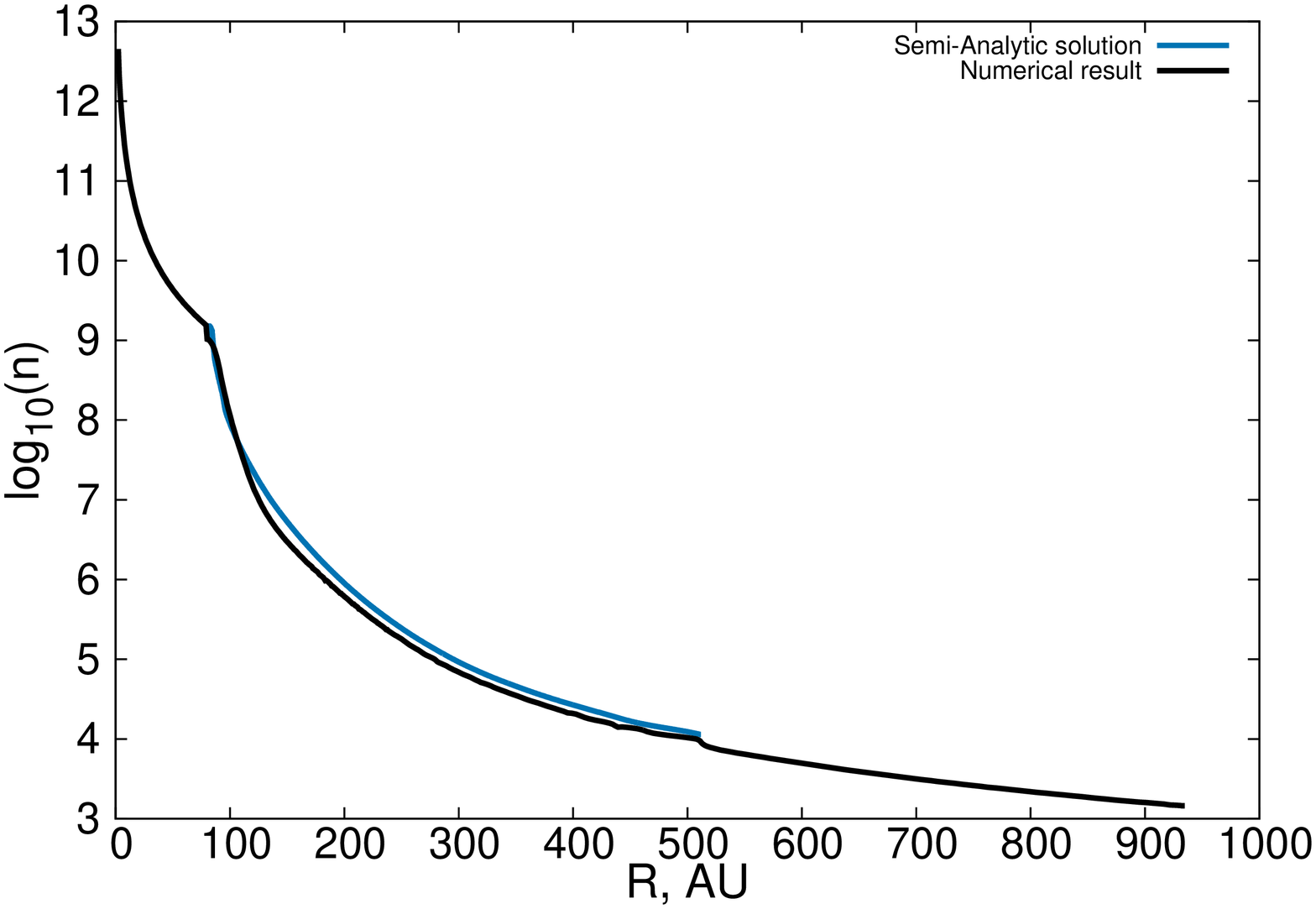}	
	\includegraphics[width=6.25cm, angle=270]{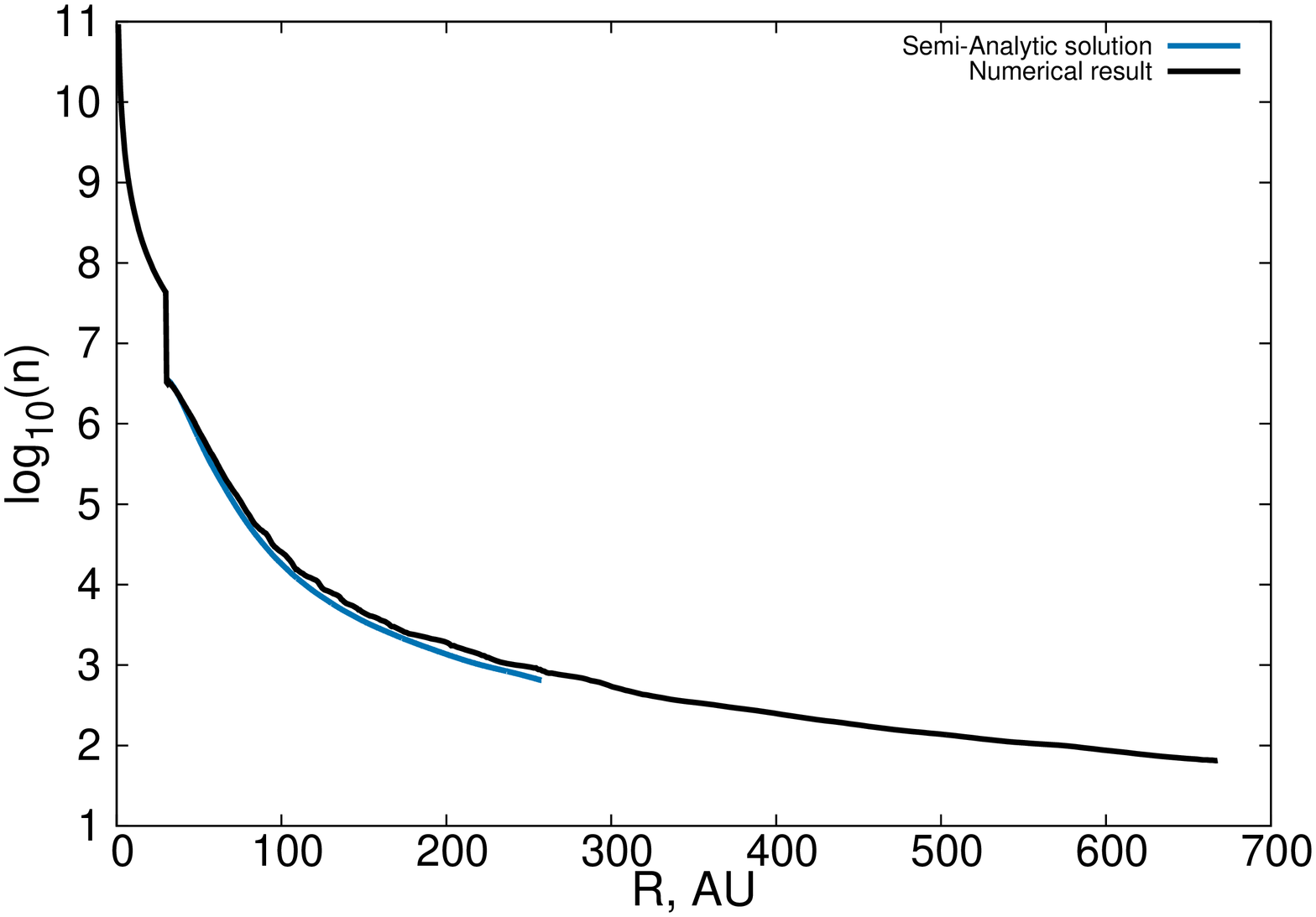}
	\caption{The final (steady state) radial density distribution for - from left to right, top to bottom - Model A, B, C and D (the spherical models).  The blue line represents the actual semi-analytic result and the black line the numerical result. In the left and right hand panels the disc is optically thick and thin respectively. The agreement is always to within a factor 2 and typically much better. }
	\label{fig:DensityPlotsSpherical}
\end{figure*}

\begin{figure*}
	\hspace{-10pt}
	\includegraphics[width=6.25cm, angle=270]{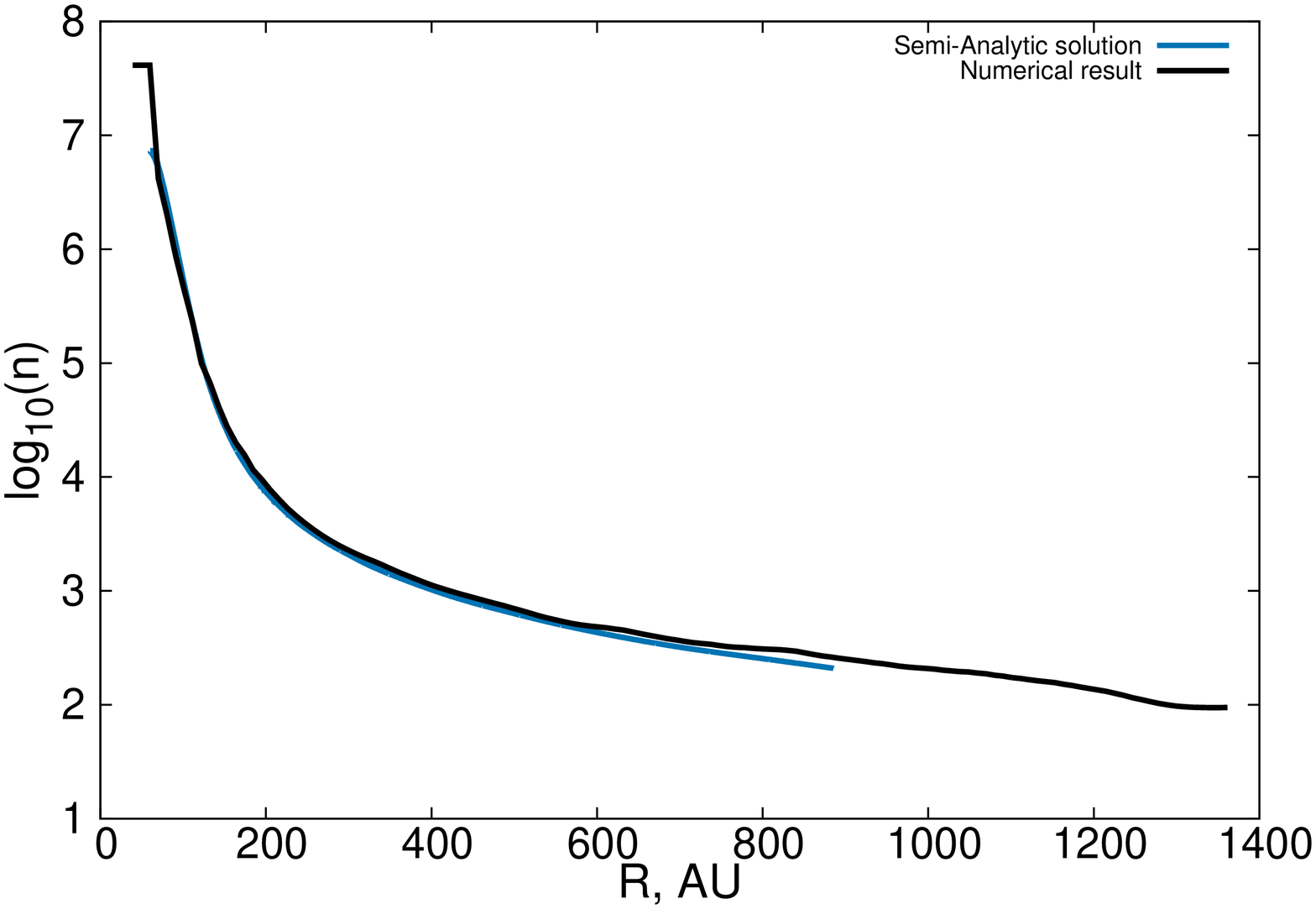}	
	\includegraphics[width=6.25cm, angle=270]{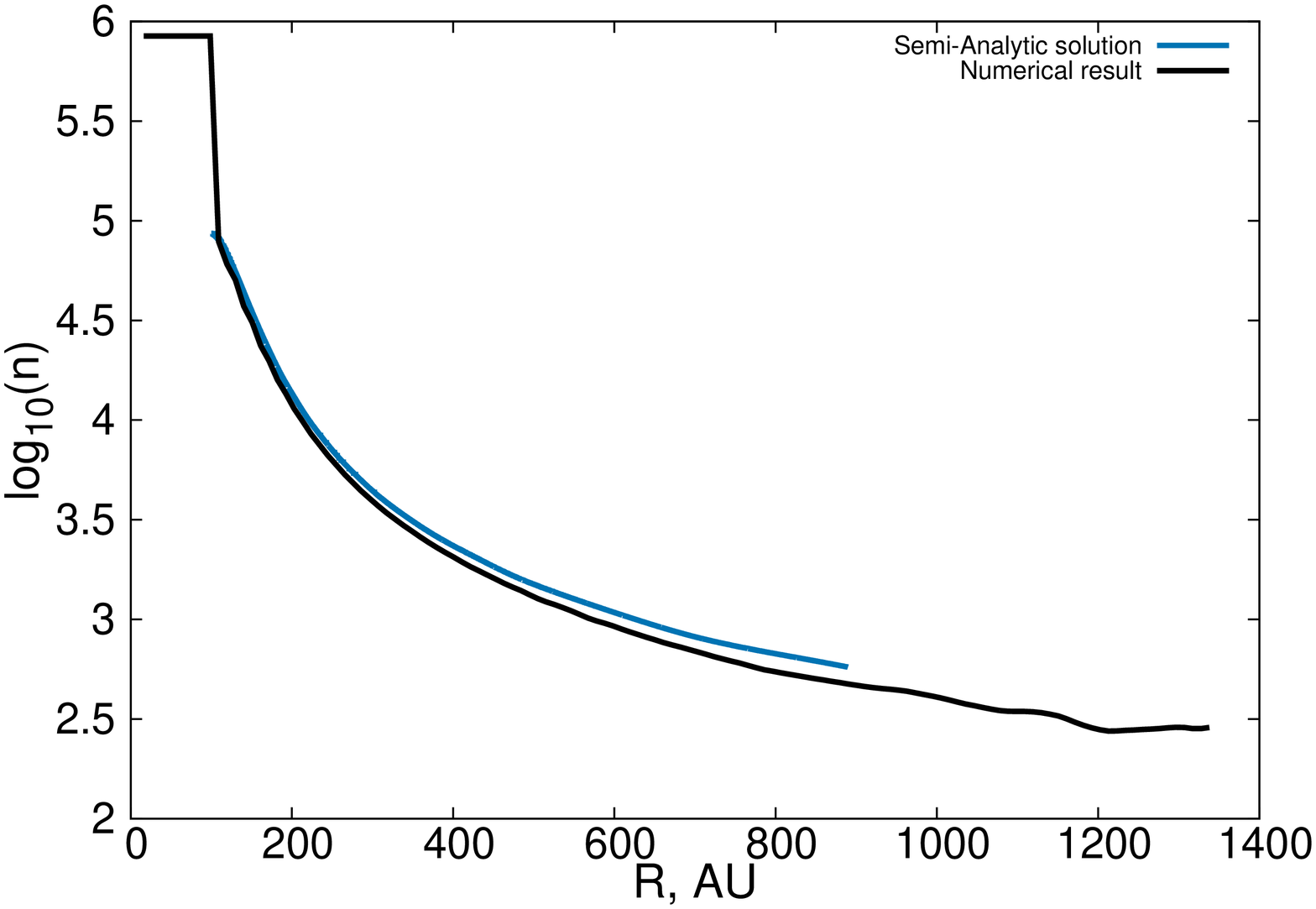}
	\caption{The final (steady state) radial density distribution for Models E and F (the cylindrical models).  The blue line represents the actual semi-analytic result. In both of these models the disc is optically thin.  The agreement is always to within a factor 2 and typically much better. }
	\label{fig:DensityPlotsCylindrical}
\end{figure*}

\subsection{Comparing steady state results with semi-analytic models}
\label{sec:comparison}
We now compare our numerical models with their corresponding semi-analytic solutions. Figures \ref{fig:DensityPlotsSpherical} and \ref{fig:DensityPlotsCylindrical} show the final steady state density distributions for the spherical and cylindrical models respectively. The blue line shows the semi--analytic result from {F16} and the black lines are our numerical results.  Both solutions are always in agreement to within a factor 2 at all radii and are typically much more accurate than this, which is excellent given that the flow density spans several orders of magnitude.  The left hand panels of Figure \ref{fig:DensityPlotsSpherical} are results in which the disc outer edge is optically thick to the incident FUV field, for all other models the disc outer edge is optically thin to the incident FUV field. 

In the optically thick cases the density distribution is continuous at the disc outer edge; however in the optically thin cases there is a contact discontinuity (there is pressure balance, but a density and temperature discontinuity). This arises because we are imposing a disc, with temperature set by the star, that is in pressure equilibrium with the comparatively hot flow. In essence we are assuming that on the inward side of the disc-flow interface the medium is optically thick, regardless of how optically thin the innermost portion of the flow is. This assumption can be probed in future dynamical models where the disc is not imposed as a boundary condition.

There is a slight numerical complication, particularly in the optically thin models where there is a contact discontinuity at the disc outer edge, in that the density is changing so rapidly that the result can be sensitive to how well the disc outer edge is captured by the simulation grid. Indeed, {F16} demonstrated that there are regimes where the flow properties (in particular the mass loss rate) are very sensitive to the disc outer radius (c.f. the right hand panel of Figure 12 in their paper). Resolution can therefore significantly modify the simulation results in some regimes by providing different effective disc outer radii. A solution to this is to impose a cell at the disc outer radius. However in future models where the disc might evolve it is not clear where this cell should be forced, making this approach not ideal.   Furthermore the steep density contrast can still give rise to slightly numerically sensitive results depending on the order of hydrodynamics scheme used, flux limiter etc. Differences between the semi--analytic and dynamical models can also arise depending on the temperature grid (and how it is interpolated between) in the semi--analytic models. There are therefore many sources of possible discrepancy between the models and semi--analytic solutions.

\subsubsection{Mass loss rates}
The mass loss rate at some point in the flow of our spherical models is computed using
\begin{equation}
	\dot{M} = 4\pi R^2 \rho \dot{R} \mathcal{F}
\end{equation}
where $\mathcal{F}$ is the fraction of solid angle subtended by the disc outer edge, which \cite{2004ApJ...611..360A} define as
\begin{equation}
	\mathcal{F} = \frac{H_d}{\sqrt{H_d^2+R_{d}^2}}. 
\end{equation}
Similarly, the mass loss rate in the cylindrical case is given by 
\begin{equation}
\dot{M}=2\pi R H_{d} \rho \dot{R}.
\end{equation}
where $H_d$ is the scale height at the disc outer edge. In a steady state, this quantity is constant throughout the flow. {A discussion of the sensitivity of the mass loss rate to disc mass, disc outer radius, FUV field strength and also to the level of grain growth in the disc (affecting the dust population entrained in the flow) is given in section 7.2 of F16 based on their large parameter space of semi--analytic models.}

\begin{figure}
    \hspace{-10pt}
    \includegraphics[width=6.33cm, angle=270]{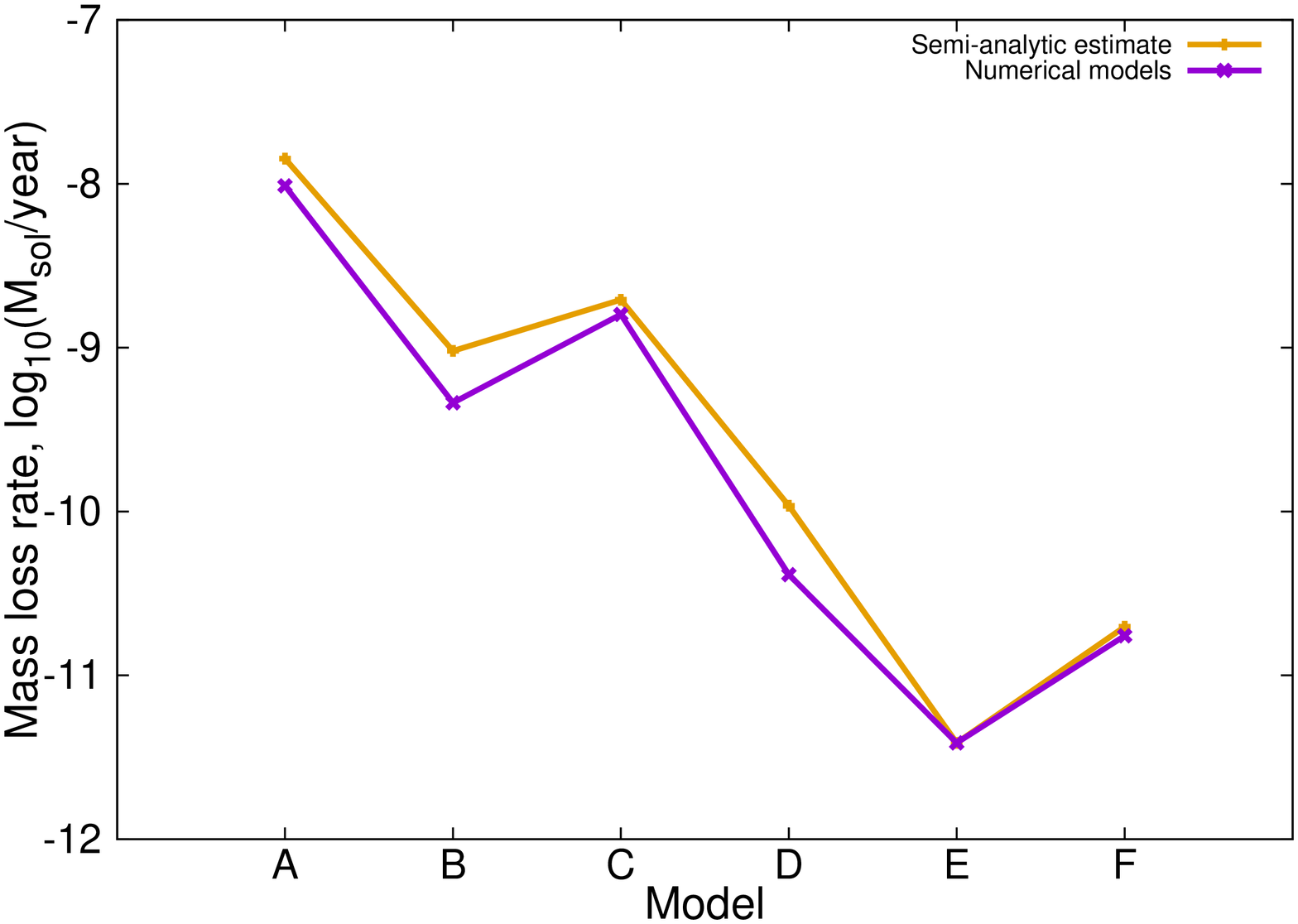}
    \caption{The mass loss rate for each model in this paper, both for the semi--analytic estimates and numerical models. Note that in the case of model E the mass loss rates are indistinguishable. }
    \label{fig:MLR}
\end{figure}

The mass loss rate for each model is shown in Figure \ref{fig:MLR}, where the orange line-points denote the mass loss rate from our numerical models and the purple that from the semi--analytic solutions. The agreement in mass loss rate is typically excellent (excluding models B and D to within 50 per cent) and model E is essentially perfect.  
Models B and D are in the regime where the mass loss rate is extremely sensitive to the disc outer edge, varying by a few orders of magnitude per $\sim10$\,AU of disc outer edge {(again, c.f. the right hand panel of Figure 12 in F16)}. Even so the mass loss rate is still only underestimated by a factor 2.5 (implying that the simulation believes there is a slightly smaller disc radius than the semi--analytic model).

\begin{figure}
	\hspace{-10pt}
	\includegraphics[width=6.34cm, angle=270]{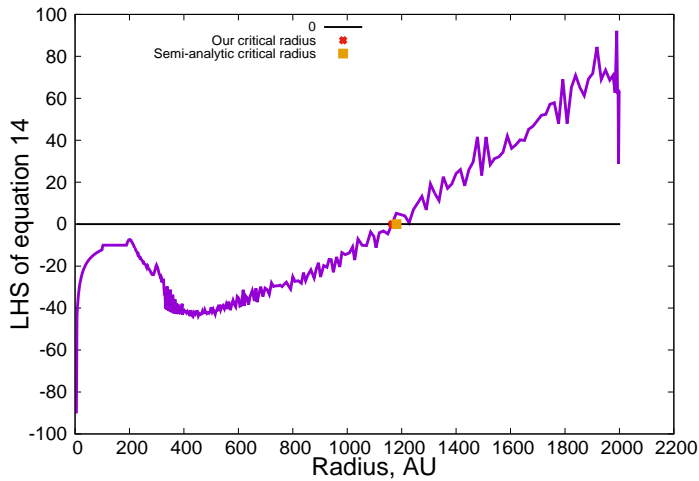}
	\caption{The radial profile of the left hand side of equation \ref{toSolve} for model A. The critical radius is the point at which this distribution crosses zero. On this plot are marked the critical radius that we compute, as well as the semi--analytic value.  }
	\label{equn14Profile}
\end{figure}

\begin{figure}
	\hspace{-10pt}
	\includegraphics[width=6.34cm, angle=270]{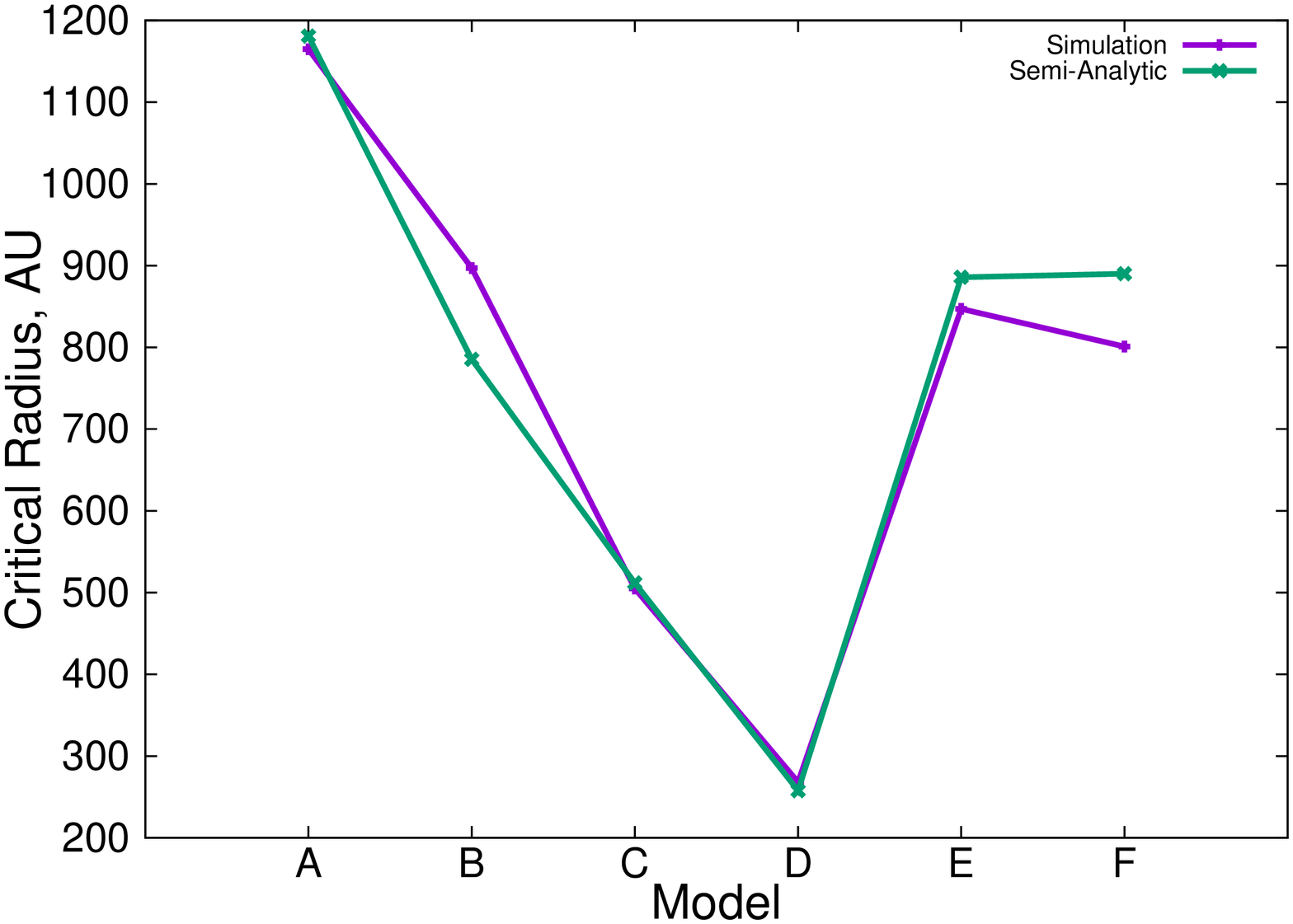}
	\caption{A comparison of the semi--analytic critical radii and approximate estimates of the critical radius derived from our simulations. }
	\label{rCritFig}
\end{figure}

\subsubsection{Critical Radii}
Following {F16} and the appendix of this paper, the critical radius is the point at which 
\begin{equation}
\label{critRad}
u^2=f+g\frac{\partial f}{\partial g},
\end{equation}
is satisfied, where $u=v/c_c$, $g=n/n_c$ and $f=T/T_c$ are the dimensionless velocity, number density and temperature, scaled to the values at the critical radius.  Multiplying through by the temperature at the critical point, we obtain that the critical radius is that at which
\begin{equation}
	\frac{v^2\mu m_{\textrm{H}}}{k_{\textrm{B}}} - T - n \frac{\textrm{d}T}{\textrm{d}n} = 0 
	\label{toSolve}
\end{equation}
is satisfied, where $v, n$ and $T$ are the local velocity, number density and temperature. The derivative in this equation is supposed to be for a given optical depth and was computed by {F16} using their temperature grids. Generally such a temperature grid will not be readily available for a code directly modelling the dynamics and PDR chemistry. 
There is an alternative, more accessible, method of computing this derivative which relies on the fact that the extinction at the critical point is small (at most $\sim10^{-2}$ for the models in this paper, see Tables \ref{modelSummaryA}, \ref{modelSummaryB}) and that the temperature is not strongly dependent on the extinction at such small values {(see Figure \ref{slabTempFig} of this paper and Figure 3 of F16)}. Under these conditions we can 
approximate the derivative spatially assuming that the optical depth doesn't vary significantly between cells. That is, if we are evaluating cell $i$, the derivative is $\left[T(i) - T(i-1)\right] / \left[n(i) - n(i-1)\right]$.

As an example, the radial profile of the left hand side of equation \ref{toSolve} is shown for model A in Figure \ref{equn14Profile}. The critical radius is the point at which this profile crosses zero. A summary of our estimated critical radii for all of our models, alongside the critical radii from the semi--analytic models, are given in Figure \ref{rCritFig}. Our agreement between the semi--analytic and numerical models in the spherical cases is very good, with the biggest discrepancy at 14 per cent in the case of model B. Models A, C and D match extremely well. In the cylindrical case the agreement is at worst within 11 per cent, in the case of model F.

\subsection{Summary of dynamical results}
We have demonstrated that the existing semi-analytic spherical (F16) and cylindrical (this paper) models are in excellent agreement with dynamical models. The agreement in the mass loss rates is also excellent, and where slight deviations arise it is in regimes where there is a good explanation as to why. We have also checked that our critical radii are consistent  with those expected from these semi--analytic models. 

A very important point is that {F16} took a different approach to \cite{2004ApJ...611..360A} in that they included non--isothermal terms in the determination of the critical radius. They found that this can lead to critical radii a factor of a few times larger than the sonic radius. They also found that the mass loss rates could sometimes be lower than \cite{2004ApJ...611..360A} predict by up to an order of magnitude (though this could, in part, be due to differences in the PDR code employed). Here, we have verified that the solutions including the non--isothermal terms are accurate when compared with consistent dynamical models. 

Finally we note that given the semi--analytic solutions are accurate, they can make useful benchmarks for future photochemical--dynamical codes (though due care is required regarding capturing of the disc outer edge, either using high resolution or forcing cells there). 

\begin{figure*}
	\hspace{-10pt}
	\includegraphics[width=6.275cm, angle=270]{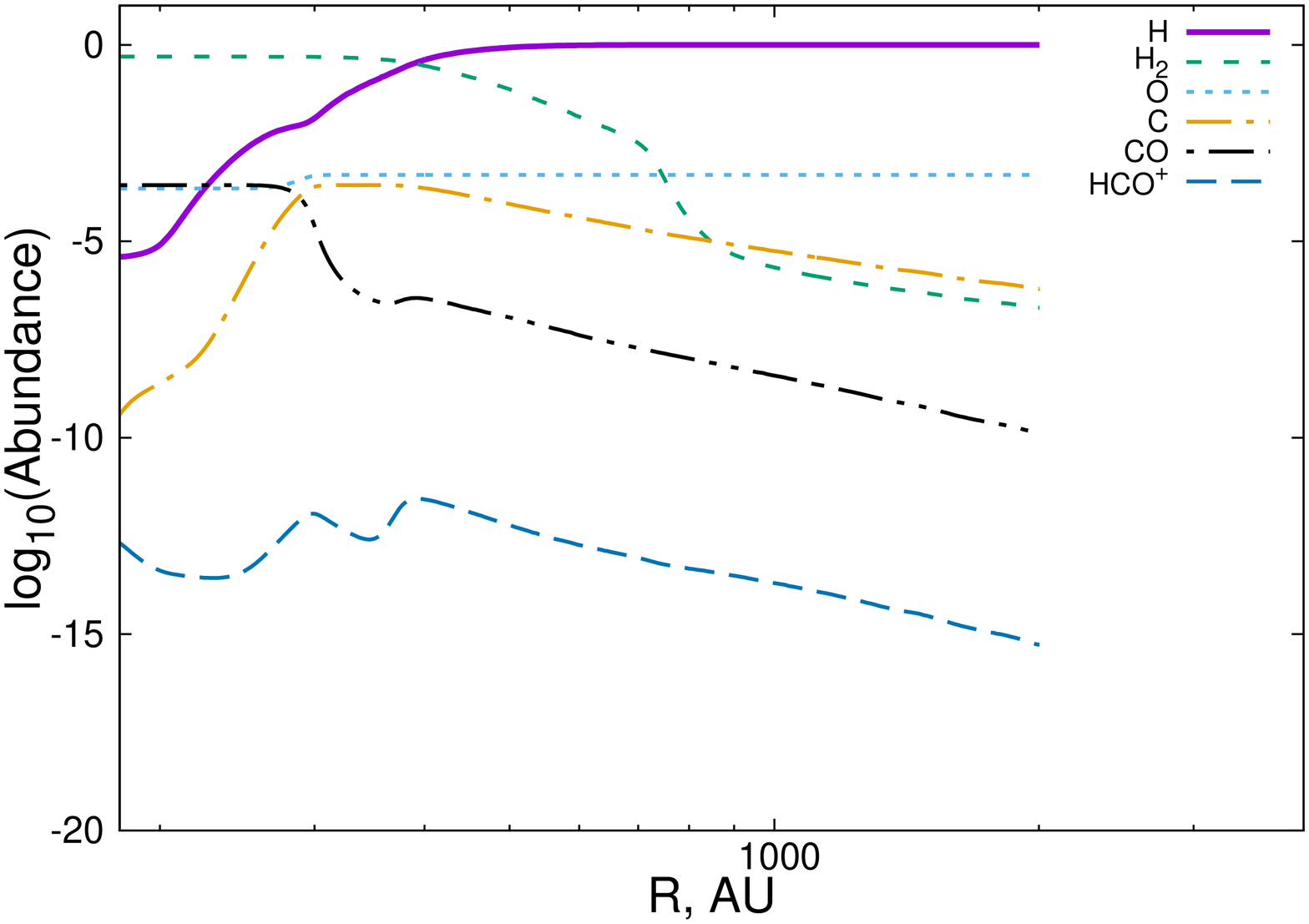}	
	\includegraphics[width=6.275cm, angle=270]{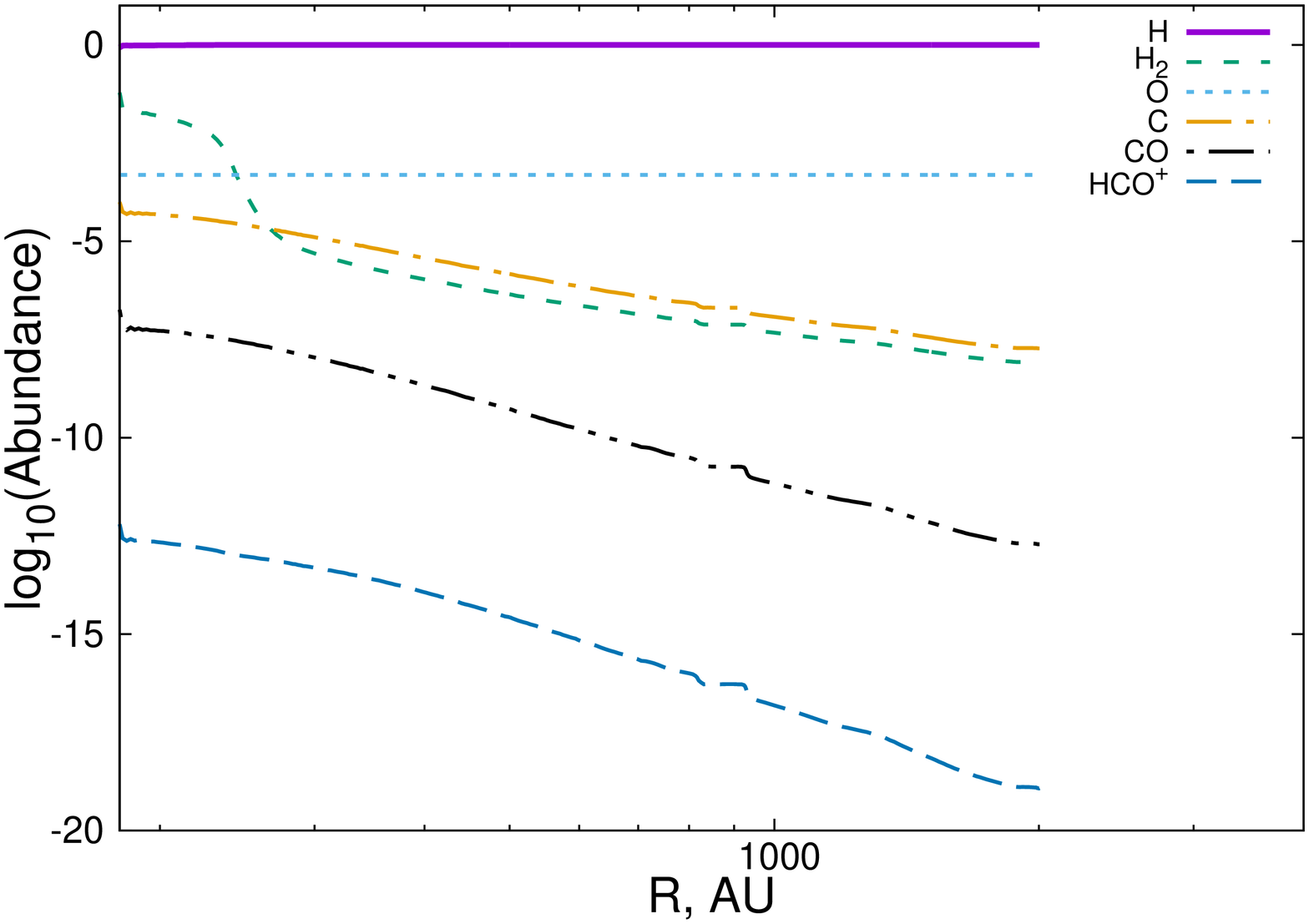}	

	\vspace{10pt}
	\hspace{-10pt}	
	\includegraphics[width=6.275cm, angle=270]{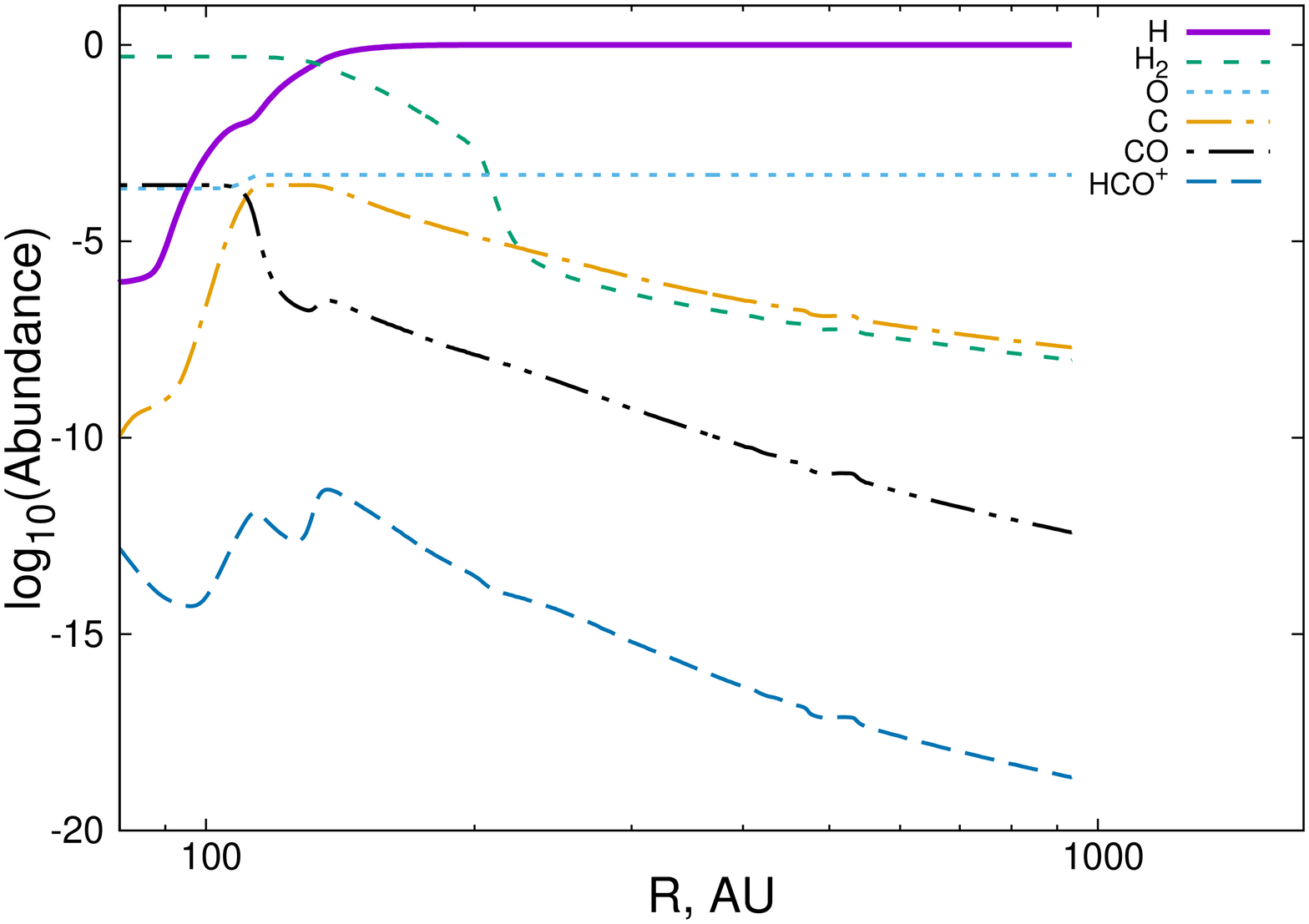}	
	\includegraphics[width=6.275cm, angle=270]{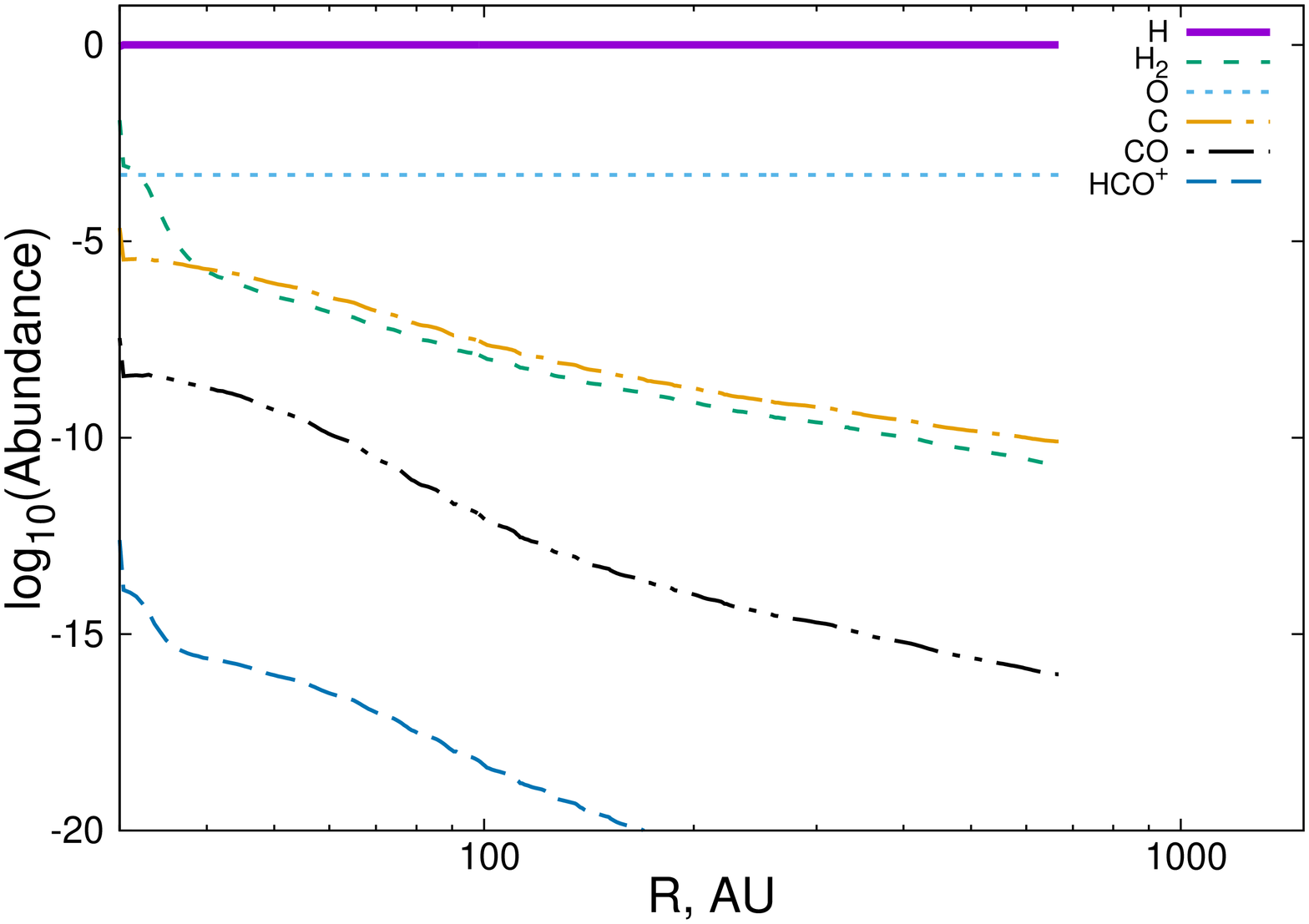}		
	\caption{The radial profile the abundances of some key species in our photochemical--dynamical models. Panels are - from left to right, top to bottom - models A, B, C and D. The left and right hand panels are models in which the disc outer edge are optically thick and thin to the incident FUV field respectively. The left hand edge of each panel is the disc outer radius.  }
	\label{MoreChemPlots}
\end{figure*}

\subsection{Flow composition and thermal structure}
\label{sec:flowComp}
Our direct modelling approach means that we automatically yield the composition of the flow. Figure \ref{MoreChemPlots} shows the radial composition profiles of some key species from our reduced network (recall that all of the species in the network are summarised in Table \ref{PDRGuts}) for our four spherical models (A--D). Again, the left hand panels are models in which the disc is optically thick to the FUV (models A, C) and the right hand panels optically thin (models B, D). The left hand edge of each panel in Figure \ref{MoreChemPlots} is the disc outer edge. We do not include the \textit{disc} composition since in our models: 
\begin{enumerate}
	\item{We do not include key gas--grain chemical processes that set the composition in the disc}
	\item{The disc temperature is imposed, rather than being set by the incident radiation field. This means that the temperature in optically thin models interior to the disc outer edge may not be consistent with what the chemical model would otherwise compute. }
\end{enumerate}
Future models in which the disc is allowed to evolve (rather than acting as a boundary condition) and in which gas--grain chemistry is included will allow for a more robust determination of the disc composition.

\begin{figure}
	\hspace{-15pt}
	\includegraphics[width=6.3cm, angle=270]{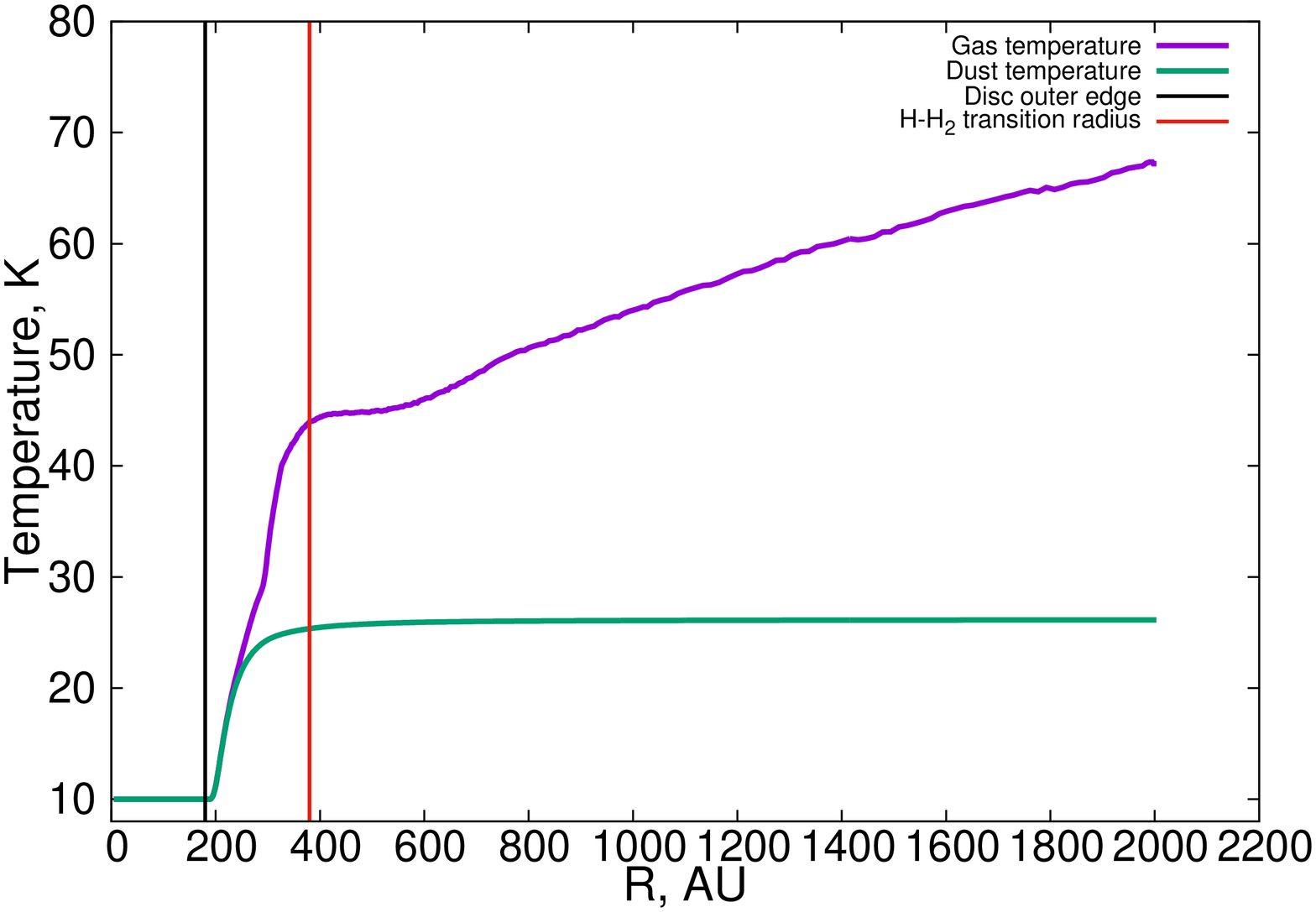}
	\caption{The dust and gas temperature profiles for model A (without the imposed disc temperature structure discussed in section \ref{bounds}). The black vertical line represents the disc outer edge and the red vertical line the location of the H--H$_2$ transition.  In model A, the disc outer edge is optically thick to the incident FUV and the H--H$_2$ transition takes place at some point in the flow (at $\sim380$\,AU). This transition marks the point at which the optical depth approaches unity, the temperature in the flow decreases substantially and a number of chemical transitions occur.}
	\label{fig:temperature}
\end{figure}

\begin{figure}
	\hspace{-15pt}
	\includegraphics[width=6.3cm, angle=270]{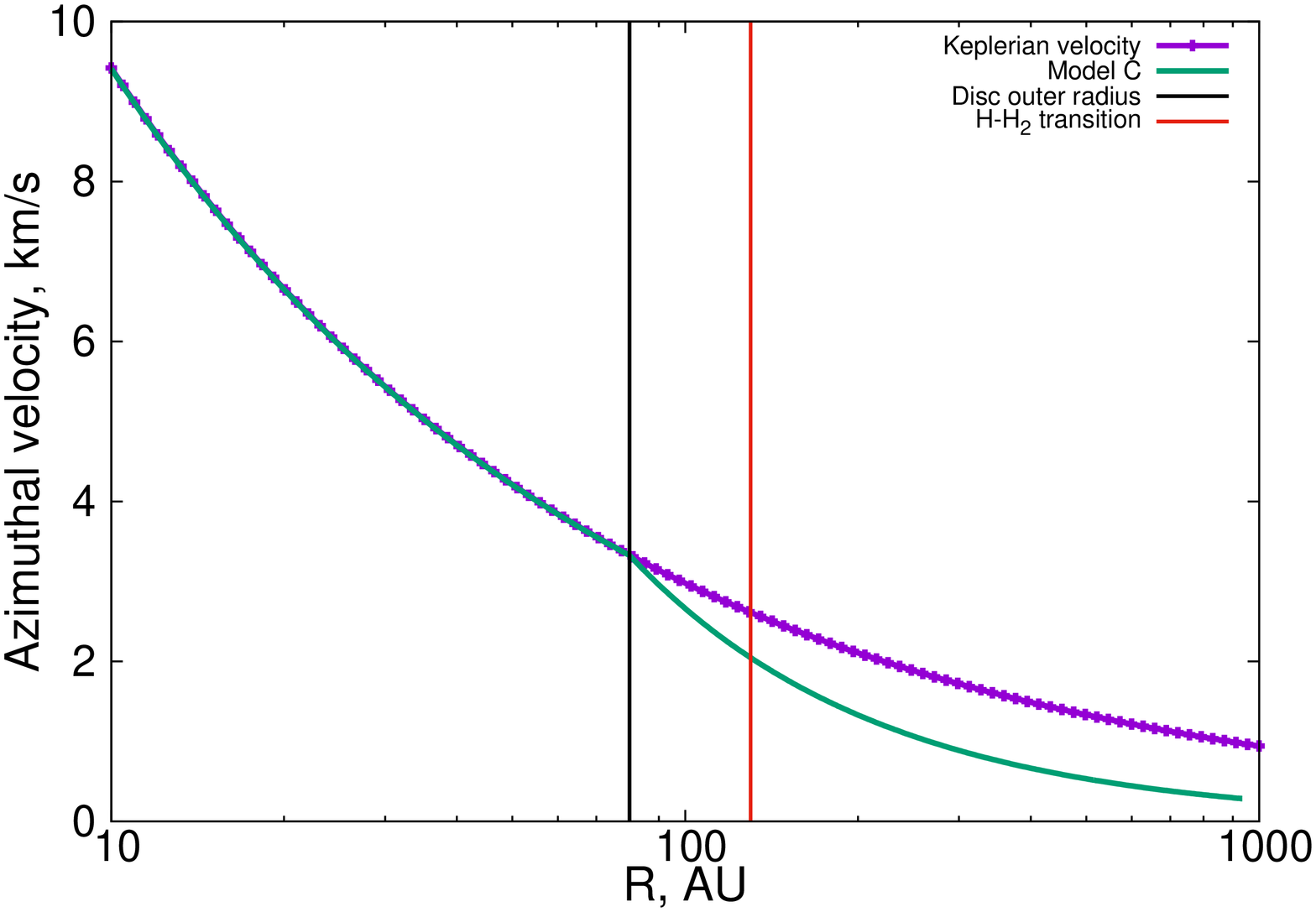}	
	\caption{A comparison of the keplerian rotation profile with that resulting from Model C. Between the disc outer edge and the H--H$_2$ transition (which is in the flow) the velocity is sub--Keplerian, but also abundant in CO, which may provide an observational diagnostic of externally photoevaporating discs. }
	\label{velocities}
\end{figure}

The optically thick models (left hand panels of Figure \ref{MoreChemPlots}) are clearly chemically distinct from the optically thin ones (right hand panels). The transition from predominantly atomic to predominantly molecular hydrogen (the H--H$_2$ transition) occurs at an optical depth of around unity. In the optically thick models this therefore takes place somewhere in the flow itself. In the optically thin regime if the transition is not exactly at the disc outer edge it will not occur at some significantly smaller radius, since the optical depth into the disc increases rapidly. The H--H$_2$ transition is significant in PDR models because it is associated with large gradients in the thermal and  chemical structure. In the case of our models, radially interior to the H--H$_2$ transition, the temperature is much lower than beyond it, which we illustrate using the temperature profile of  models A  in Figure \ref{fig:temperature}. For model A, exterior to the H--H$_2$ transition, the temperature only decreases by about 20\,K over $\sim1800$\,AU throughout the rest of the modelled flow ($\sim0.01$\,K\,AU$^{-1}$). Conversely interior to the transition, from that point to the disc outer edge, the temperature decreases by over 30\,K in only 200\,AU ($0.15$\,K\,AU$^{-1}$, over an order of magnitude faster). Within this component of the flow interior to the H--H$_2$ transition the gas and dust also eventually become thermally coupled. Chemically the abundances of CO, C and HCO$^{+}$ - key observational tracers - are all sensitive to this optical depth of around unity. 

With \textsc{torus-3dpdr} our simulations also yield the level populations and line emissivities of the various species in the calculation. However, because our models in this paper all use the radially dominated photon escape approximation (i.e. optical depths along directions not parallel to the disc mid--plane are infinite) and assume spherical symmetry to be consistent with the semi--analytic models, they could be misleading when compared too directly with real observations. We do note, however, that in the optically thick regime there is a significant abundance of CO in the flow between the disc outer edge and the H--H$_2$ transition. The azimuthal velocity of this material is sub--Keplerian, which we illustrate in Figure \ref{velocities}. It is therefore possible that sub--Keplerian CO emission towards the outer regions of protoplanetary discs could be an indicator of externally irradiated discs {(a possibility also suggested by F16)}. Future models more directly translatable to observations are required to confirm whether this should be observable.

\section{Summary and conclusions}
We present photochemical--dynamical simulations of protoplanetary discs that are externally irradiated by  FUV radiation fields. Such models are novel, because in order to compute the temperature structure in the flow that is driven off of the disc, one has to solve for the photon dominated region chemistry, which we do ``on the fly'' in our numerical models. In this first paper, we aim to verify that previous semi--analytic models of this process are in agreement with our numerical results. We draw the following main conclusions from this work. \\

\noindent 1. We find excellent agreement between our numerical models and semi--analytical solutions for the flow structure, both for the spherical scenario considered by {F16} and also a new cylindrical geometry presented in this paper. Despite the flow spanning many orders of magnitude in density, sometimes over only $\sim500$\,AU, we find agreement in the wind density is always to within a factor 2 at all radii (and typically much better than this). We also obtain critical radii and mass loss rates consistent (at worst to within 14 per cent and a factor 2.5 respectively) with those predicted by the semi--analytic models. The level of agreement is consistent with
what is expected given the sensitivity of the flow rates to
the disc outer edge (see 3.)\\

\noindent 2. The models of {F16} (with which our simulations agree) differ from past models by  \cite{2004ApJ...611..360A} in that they included non--isothermal terms when determining the location of the critical radius in the flow. These terms lead to larger critical radii by up to a factor of a few and, in some scenarios, lower mass loss rates by up to an order of magnitude. We therefore verify that the non--isothermal terms are important for determining semi--analytic solutions and confirm the accuracy of the mass flow rates
(given the assumptions in our PDR model). \\

\noindent 3. The flow properties can be very sensitive to the exact disc outer radius. A source of discrepancy between numerical models and semi--analytic solutions in such a regime can therefore be how well the disc outer edge is captured by the simulation, particularly in terms of resolution. \\

\noindent 4. Our models also yield information on the composition of discs irradiated by an  external FUV field (with the caveats that we cannot compute accurate compositions in the disc itself and our models are 1 dimensional). There is a thermal and chemical distinction between models in which the wind flow is optically
thin to the incident FUV field throughout its radial extent as opposed to
cases where the flow becomes optically thick to the FUV exterior to the
the disc edge.  Of particular interest is that in the latter case there is a healthy abundance of CO between the disc outer edge and the H--H$_2$ transition, which has a rotational velocity that is sub--Keplerian. If a more dedicated study confirms this signature, it could possibly provide a useful diagnostic of externally irradiated discs.

\section*{Acknowledgements}
{We thank the anonymous referee for their comments on the manuscript}. During most of this work TJH was funded by the STFC consolidated grant ST/K000985/1 at the University of Cambridge and is now funded by the Imperial College London Junior Research Fellowship scheme. DB thanks the STFC for a studentship. This work has been supported by the DISCSIM project, grant agreement 341137 funded by the European Research Council under ERC-2013-ADG. This work was undertaken on the COSMOS Shared Memory system at DAMTP, University of Cambridge operated on behalf of the STFC DiRAC HPC Facility. This equipment is funded by BIS National E-infrastructure capital grant ST/J005673/1 and STFC grants ST/H008586/1, ST/K00333X/1.

\bsp

\bibliographystyle{mnras}
\bibliography{molecular}


\appendix

\section{Cylindrical semi-analytic flow solution}
\label{cylsol}

\begin{figure}
    \hspace{-10pt}
    \includegraphics[width=6.32cm,angle=270]{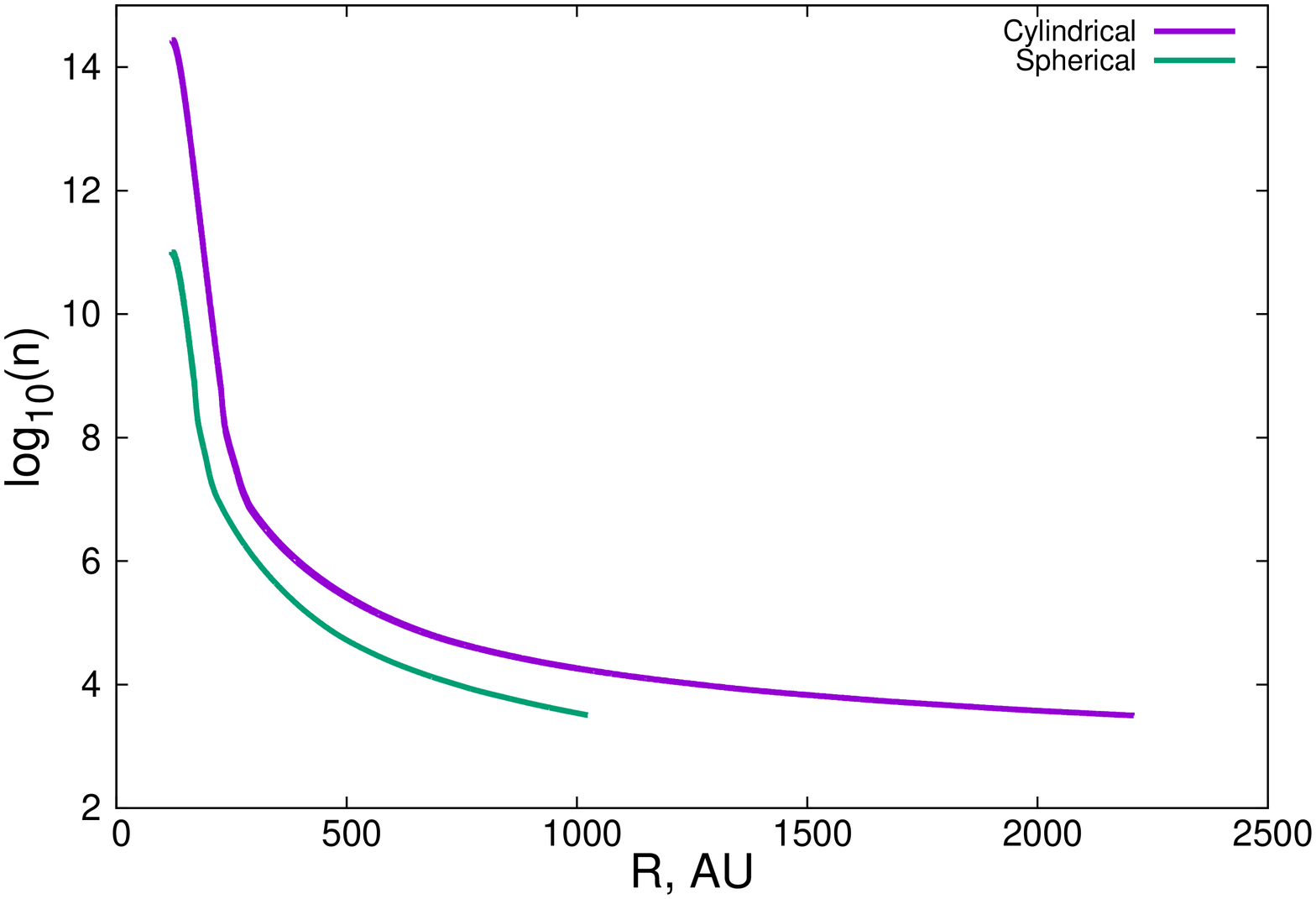}
    
    \hspace{-10pt}
    \includegraphics[width=6.32cm,angle=270]{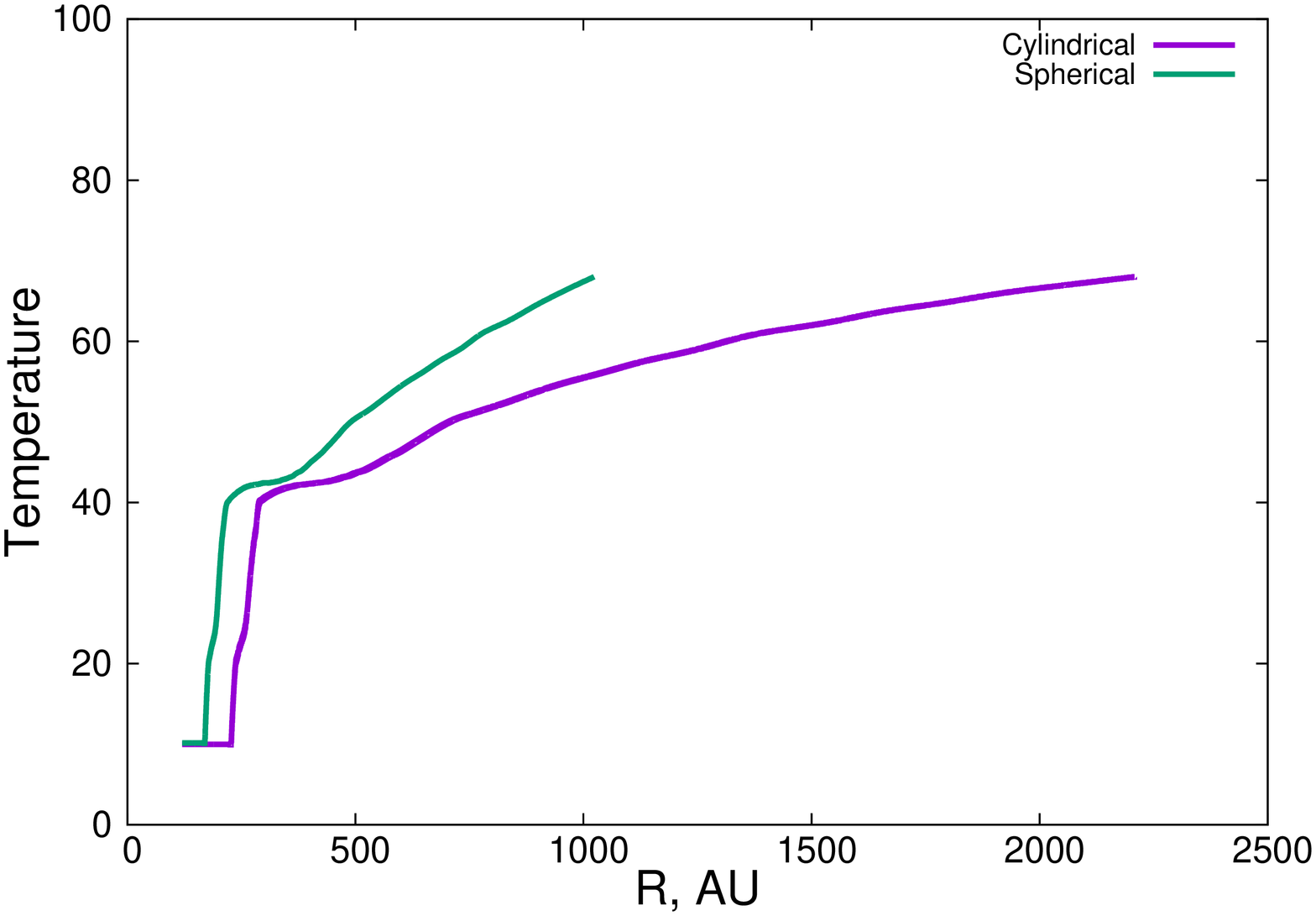}  
    \caption{A comparison of the spherical and cylindrical semi-analytic solutions for a disc with $T_{\text{c}}=68 \; \text{K}$ and $R_{\text{d}}=120 \; \text{AU}$.}
    \label{sph_vs_cyl}
\end{figure}

For completeness, we present the extension of the semi-analytic solution from {F16} to the 2D cylindrical case. The derivation is given in reference to {F16} to aid comparison with the spherical solution, and only those equations which are substantially altered are shown here.

The geometry in the 2D cylindrical case is an axisymmetric flow with the maximum vertical extent of the flow above the mid-plane $z$ being small compared to the cylindrical radius $R$. Thus we solve the fluid equations in the plane and assume that this solution is applicable over the relevant range of heights above the plane.
The gravitational term in the momentum equation should be $F(R,z)=-GM/(R+z)^2$, but since typically $z\leq0.1R$ we make the simplifying assumption that $F(R)=-GM/R^2$. Under this prescription the momentum equation is unchanged in the transform from spherical radius $r$ to cylindrical radius $R$, however the cylindrical geometry modifies the time-independent continuity equation to
\begin{equation}
\label{aeq:continuity}
\dot{M}=2\pi R H_{\text{d}} \mu_{\text{H}} n v,
\end{equation}
where $H_{\text{d}}$ is the height of the disc. Making the dimensionless transformations $\xi \equiv R/R_{\text{d}}$, $f\equiv T/T_{\text{c}}$, $g \equiv n/n_{\text{c}}$ and $u \equiv v/c_{\text{s,c}}$ the continuity equation becomes:
\begin{equation}
\label{aeq:dlesscont}
\xi gu=C,
\end{equation}
where
\begin{equation}
\label{aeq:C}
C=\frac{\dot{M}}{2\pi R_{\text{d}}H_{\text{d}}\mu m_{\text{H}}n_{\text{c}}c_{\text{s,c}}}=\frac{R_{\text{c}}}{R_{\text{d}}}\frac{v_{\text{c}}}{c_{\text{s,c}}}.
\end{equation}
Using the continuity equation to isolate the dimensionless density $g$ in the momentum equation we obtain:
\begin{equation}
\label{aeq:dlessmoment}
\frac{d\ln u}{d\xi}(u^2-f-g\frac{\partial f}{\partial g})=\frac{1}{\xi}(f+g\frac{\partial f}{\partial g})-\beta \frac{\xi-1}{\xi^3}+\tau_{\text{d}}g\frac{\partial f}{\partial \tau}.
\end{equation}
Procceeding in parallel to {F16}, we define the critical point as occurring when both sides of Equation \ref{aeq:dlessmoment} are zero. This implicitly defines a critical velocity:
\begin{equation}
\label{aeq:u2}
u_{\text{c}}^2=f+g\frac{\partial f}{\partial g},
\end{equation}
which is related to the sound speed $u_{\text{s}}^2=f$ by an additional term which accounts for departures from isothermality. Evaluating equation \ref{aeq:dlessmoment} at the dimensionless critical radius $\xi_{\text{c}}=R_{\text{c}}/R_{\text{d}}$ we obtain:
\begin{equation}
\label{aeq:criticalpoint}
g\tau_{\text{d}}\frac{\partial f}{\partial \tau}\xi_{\text{c}}^3+u_{\text{c}}^2\xi_{\text{c}}^2-\beta \xi_{\text{c}}+\beta=0.
\end{equation}

We find the sonic radius $\xi_{\text{s}}$, which we require as an initial estimate of the critical radius $\xi_{\text{c}}$, using the isothermal limit of equation \ref{aeq:criticalpoint}:
\begin{equation}
\label{aeq:sonicradius}
\xi_{\text{s}}=\frac{\beta}{2}\left[1+\left(1-\frac{4}{\beta}\right)^{1/2}\right],
\end{equation}
and note that $\beta$ is fixed for a given $T_{\text{c}}$. This definition requires that $\beta>4$.

Calculating the location of the critical radius $\xi_{\mathrm{c}}$ requires an assumption about the density structure of the flow beyond the critical radius. We cannot assume that the velocity is approximately constant as in spherical symmetry, since this leads to an infinite mass in the wind, so we instead assume $u =u_{\text{c}} f(R/R_{\text{c}})$ for $R>R_{\text{c}}$ where $f(x)$ is an increasing function with $f(1)=1$. Then the continuity equation gives:
\begin{equation}
\label{aeq:boundarydensity}
n(R)=\frac{n_{\text{c}}}{(\frac{R}{R_{\text{c}}})f(\frac{R}{R_{\text{c}}})}, \text{ for } R>R_{\text{c}},
\end{equation}
and consequently 
\begin{equation}
\label{aeq:boundarytau}
\tau_{\text{c}}=\sigma_{\text{FUV}}\int_{R_{\text{c}}}^{\infty}\frac{n_{\text{c}}}{(\frac{R'}{R_{\text{c}}})f(\frac{R'}{R_{\text{c}}})}dR'=\alpha \sigma_{\text{FUV}} R_{\text{c}}n_{\text{c}},
\end{equation}
where:
\begin{equation}
\label{aeq:alpha}
\alpha=\int_1^{\infty}\frac{dx}{xf(x)}
\end{equation}
is a constant factor that parametrizes the velocity profile beyond the critical radius. In practise the solutions are only weakly dependent on this boundary condition so we fixed $\alpha=1$, which corresponds to $f(x)=x$, ie. a linearly increasing wind velocity beyond the critical radius.

The form of the remainder of the derivation is identical to  {F16} save for the propagation of the altered coefficients. For convenience we give the form of the coefficients used in the expansion about the critical point,
\begin{equation}
\label{aeq:expansion}
\frac{\delta u}{\delta \xi}\Bigr|_{\xi_{\text{c}}}=\frac{-B+\sqrt{B^2-4AD}}{2A},
\end{equation}
where:
\begin{equation}
\label{aeq:expansionA}
A=2u+\frac{2g}{u}\frac{\partial f}{\partial g}+\frac{g^2}{u}\frac{\partial^2 f}{\partial g^2};
\end{equation}
\begin{equation}
\label{aeq:expansionB}
B=2\tau_{\text{d}}g\frac{\partial f}{\partial \tau}+\frac{4g}{\xi}\frac{\partial f}{\partial g} +\frac{2g^2}{\xi}\frac{\partial^2 f}{\partial g^2}+2\tau_{\text{d}}g^2\frac{\partial^2 f}{\partial \tau \partial g};
\end{equation}
\begin{align}
\label{aeq:expansionD}
D=u\times&\left( \frac{u^2}{\xi^2}-\beta \frac{2\xi-3}{\xi^4} +\frac{2g}{\xi^2}\frac{\partial f}{\partial g}+\frac{2\tau_{\text{d}} g}{\xi}\frac{\partial f}{\partial \tau} \right.\nonumber \\
                  & \left. + \frac{g^2}{\xi^2}\frac{\partial^2 f}{\partial g^2} +\frac{2\tau_{\text{d}} g^2}{\xi}\frac{\partial ^2 f}{\partial \tau \partial g}+\tau_{\text{d}}^2g^2\frac{\partial ^2 f}{\partial \tau^2}  \vphantom{\frac12}\right) ,
\end{align}
and all the quantities are evaluated at $\xi=\xi_{\text{c}}$. 

A comparison of the density and temperature profile for identical discs in spherical and cylindrical geometries is given in Figure \ref{sph_vs_cyl}. The difference in geometry can induce an inner flow density almost 3 orders of magnitude higher at the disc outer edge and a critical radius a factor of 2 larger in the cylindrical case. 

Comparing the sonic radius obtained with cylindrical geometry $\xi_{\text{s,cyl}}$ in Equation \ref{aeq:sonicradius} to the sonic radius in spherical geometry $\xi_{\text{s,sph}}$ from {F16}, we see that for equal critical temperature $T_c$ and disk radius $R_d$ (hence equal $\beta$), $\xi_{\text{s,cyl}}/\xi_{\text{s,sph}} \approx 2+2/\beta$. Thus we would expect that the location of the cylindrical critical radius will be approximately twice that of the spherical critical radius.

\label{lastpage}
\end{document}